\documentclass{acm_proc_article-sp}
\usepackage{graphicx}
\usepackage{balance}  
\usepackage[ruled]{algorithm2e}
\usepackage[noend]{algpseudocode}
\makeatletter
\renewcommand{\@algocf@capt@plain}{above}
\makeatother
\usepackage{pgfplots}
\usepackage{times,amsmath,epsfig}
\usepackage{colortbl}
\usepackage{amssymb}
\usepackage{latexsym}
\usepackage{paralist}
\usepackage{url}
\usepackage{multicol}

\usepackage{subcaption}
\newtheorem{lem} {Lemma} 
\newcommand {\BL} {\begin{lem}} 
\newcommand {\EL} {\end{lem}} 

\newtheorem{thm} {Theorem} 
\newcommand {\BT} {\begin{thm}}
\newcommand {\ET} {\end{thm}}

\newtheorem{prob} {Problem} 
\newcommand {\BPR} {\begin{prob}}
\newcommand {\EPR} {\end{prob}}

\newtheorem{defi} {Definition} 
\newcommand {\BDE} {\begin{defi}}
\newcommand {\EDE} {\end{defi}}

\newcommand{\D}{{\cal D}}

\begin{document}


\title{Probabilistic Threshold Indexing \\ for Uncertain Strings}



%
%
%
%

\numberofauthors{4} 

\author{
%
%
\alignauthor
Sharma Thankachan\\
       \affaddr{Georgia Institute of Technology}\\
       \affaddr{Georgia, USA}\\
       \email{thanks@csc.lsu.edu}
\alignauthor
Manish Patil\\
       \affaddr{Louisiana State University}\\
       \affaddr{Louisiana, USA}\\
       \email{mpatil@csc.lsu.edu}
\alignauthor
Rahul Shah\\
       \affaddr{Louisiana State University}\\
       \affaddr{Louisiana, USA}\\
       \email{rahul@csc.lsu.edu}     
\and 
\alignauthor
Sudip Biswas\\
       \affaddr{Louisiana State University}\\
       \affaddr{Louisiana, USA}\\
       \email{sbiswa7@csc.lsu.edu}
}

\maketitle

\begin{abstract}
Strings form a fundamental data type in computer systems. String searching has been extensively studied since the inception of computer science. Increasingly many applications have to deal with imprecise strings or strings with fuzzy information in them. String matching becomes a probabilistic event when a string contains uncertainty, i.e. each position of the string can have different probable characters with associated probability of occurrence for each character. Such uncertain strings are prevalent in various applications such as biological sequence data, event monitoring and automatic ECG annotations. We explore the problem of indexing uncertain strings to support efficient string searching.
In this paper we consider two basic problems of string searching, namely substring searching and string listing. In substring searching, the task is to find the occurrences of a deterministic string in an uncertain string. We formulate the string listing problem for uncertain strings, where the objective is to output all the strings from a collection of strings, that contain probable occurrence of a deterministic query string. Indexing solution for both these problems are significantly more challenging for uncertain strings than for deterministic strings. Given a construction time probability value $\tau$, our indexes can be constructed in linear space and supports queries in near optimal time for arbitrary values of probability threshold parameter greater than $\tau$. To the best of our knowledge, this is the first indexing solution for searching in uncertain strings that achieves strong theoretical bound and supports arbitrary values of probability threshold parameter. We also propose an approximate substring search index that can answer substring search queries with an additive error in optimal time. We conduct experiments to evaluate the performance of our indexes.
\end{abstract}

\section{Introduction}
\label{intro}
String indexing has been one of the key areas of computer science. Algorithms and data structures of string searching finds application in web searching, computational biology, natural language processing, cyber security, etc. The classical problem of string indexing is to preprocess a string such that query substring can be searched efficiently. Linear space data structures are known for this problem which can answer such queries in optimal $O(m+occ)$ time, where $m$ is the substring length and $occ$ is the number of occurrences reported.

Growth of the internet, digital libraries, large genomic projects have contributed to enormous growth of data. As a consequence, noisy and uncertain data has become more prevalent. Uncertain data naturally arises in almost all applications due to unreliability of source, imprecise measurement, data loss, and artificial noise. For example sequence data in bioinformatics is often uncertain and probabilistic. Sensor networks and satellites inherently gather noisy information. 

Existing research has focused mainly on the study of regular or deterministic string indexing. In this paper we explore the problem of indexing uncertain strings. We begin by describing the uncertain string model, possible world semantics and challenges of searching in uncertain strings.

Current literature models uncertain strings in two different ways: the string level model and the character level model. In string level model, we look at the probabilities and enumerate at whole string level, whereas character level model represents each position as a set of characters with associated probabilities. We focus on the character level model which arises more frequently in applications. Let $S$ be an uncertain string of length $n$. Each character $c$ at position $i$ of $S$ has an associated probability $pr(c^i)$. Probabilities at different positions may or may not contain correlation among them. Figure 1(a) shows an uncertain string $S$ of length $5$. Note that, the length of an uncertain string is the total number of positions in the string, which can be less than the total number of possible characters in the string. For example, in Figure 1(a), total number of characters in string $s$ with nonzero probability is $9$, but the total number of positions or string length is only $5$.

"Possible world semantics" is a way to enumerate all the possible deterministic strings from an uncertain string. Based on possible world semantics, an uncertain string $S$ of length $n$ can generate a deterministic string $w$ by choosing one possible character from each position and concatenating them in order. We call $w$ as one of the possible world for $S$. Probability of occurrence of $w=w_1w_2\dots w_n$ is the partial product $pr(w_1^1) \times pr(w_2^{2}) \times\dots \times pr(w_n^n)$. The number of possible worlds for $S$ increases exponentially with $n$. Figure 1(b) illustrates all the possible worlds for the uncertain string $S$ with their associated probability of occurrence.

A meaningful way of considering only a fraction of the possible worlds is based on a probability threshold value $\tau$. We consider a generated deterministic string $w=w_1w_2\dots w_n$ as a valid occurrence with respect to $\tau$, only if it has probability of occurrence more than $\tau$. The probability threshold $\tau$ effectively removes lower probability strings from consideration. Thus $\tau$ plays an important role to avoid the exponential blowup of the number of generated deterministic strings under consideration.

\begin{figure}[h!]
\begin{center}
\tiny
\begin{subfigure}{0.4\textwidth}
\begin{center}
    \begin{tabular}{ | c | c | c | c | c | c |}
    \hline
    Character & S[1] & S[2] & S[3] & S[4] & S[5]\\ \hline
    a & .3 & .6 & 0 &.5 & 1 \\ \hline
    b & .4 &  0 & 0 & 0 & 0 \\ \hline
    c &  0 & .4 & 0 &.5 & 0 \\ \hline
    d & .3 &  0 & 1 & 0 & 0 \\ \hline
    \end{tabular}
    \caption{Uncertain string $S$}
\end{center}
\end{subfigure}

\begin{subfigure}{0.4\textwidth}
    \begin{tabular}{ | c | c | c | c | c | c |}
    \hline
    w & Prob(w) & w & Prob(w) & w & Prob(w)\\ \hline
    w[1] aadaa & .09 & w[5] badaa & .12 & w[9] dadaa & .09  \\ \hline
    w[2] aadca & .09 & w[6] badca & .12 & w[10] dadca & .09 \\ \hline
    w[3] acdaa & .06 & w[7] badca & .08 & w[11] dcdaa & .06 \\ \hline
    w[4] acdca & .06 & w[8] badca & .08 & w[12] dcdca & .06 \\ \hline
    \end{tabular}
        \caption{Possible worlds of $S$}
\end{subfigure}
    \caption{An uncertain string $S$ of length $5$ and its all possible worlds with probabilities.}
\end{center}
\label{figuncert} 
\end{figure}

Given an uncertain string $S$ and a deterministic query substring $p=p_1\dots p_m$, we say that $p$ matched at position $i$ of $S$ with respect to threshold $\tau$ if $pr(p_1^i)\times \dots \times pr(p_m^{i+m-1})\geq \tau$. Note that, $O(m+occ)$ is the theoretical lower bound for substring searching where $m$ is the substring length and $occ$ is the number of occurrence reported.

\subsection{Formal Problem Definition}
Our goal is to develop efficient indexing solution for searching in uncertain strings. In this paper, we discuss two basic uncertain string searching problems which are formally defined below.

\BPR [Substring Searching]
\label{def:substring}
Given an uncertain string $S$ of length $n$, our task is to index $S$ so that when a deterministic substring $p$ and a probability threshold $\tau$ come as a query, report all the starting positions of $S$ where $p$ is matched with probability of occurrence greater than $\tau$.
\EPR

\BPR [Uncertain String Listing]
\label{def:retrieval}
Let $\D$=$\{d_1$,$\dots$,$d_D\}$ be a collection of $D$ uncertain strings of $n$ positions in total.
Our task is to index $\D$ so that when a deterministic substring $p$ and a probability threshold $\tau$ come as a query, report all the strings $d_j$ such that $d_j$ contains atleast one occurrence of $p$ with probability of occurrence greater than $\tau$.
\EPR

Note that the string listing problem can be naively solved by running substring searching query in each of the uncertain string from the collection. However, this naive approach will take $O(\sum \limits_{d_i\in \D}$search time on $d_i)$ time which can be very inefficient if the actual number of documents containing the substring is small. Figure~\ref{fig_stringlisting} illustrates an example for string listing. In this example, only the string $d_1$ contains query substring "BF" with probability of occurrence greater than query threshold $0.1$. Ideally, the query time should be proportionate to the actual number of documents reported as output.
Uncertain strings considered in both these problems can contain correlation among string positions.

\begin{figure}[h!]
\begin{center}
       \begin{tabular}{c}

    \normalsize String collection $\D=\{d_1,d_2,d_3\}$:
    \end{tabular}
	\tiny
    \begin{tabular}{|ccc|ccc|ccc|}
    \hline
    $d_1[1]$ &$d_1[2]$ &$d_1[3]$ &$d_2[1]$ &$d_2[2]$ &$d_2[3]$ &$d_3[1]$ &$d_3[2]$ &$d_3[3]$\\
    A .4     &B .3     &F .5     &A .6     &B .5     &B .4     &A .4     &I .3     &A 1     \\ 
    B .3     &L .3     &J .5     &C .4     &F .3     &C .3     &F .4     &L .3     &        \\ 
    F .3     &F .3     &         &         &J .2     &E .2     &P .2     &P .3     &        \\
    	     &J .1     &         &         &         &F .1     &         &T .3     &        \\
    	
    \hline
    \end{tabular}
   
       \begin{tabular}{c}

    \normalsize Output of string listing query $("BF",0.1)$ on $\D$ is: $d_1$
    \end{tabular}
\end{center}
 \caption{String listing from an uncertain string collection $\D=\{d_1,d_2,d_3\}$.}
\label{fig_stringlisting}
\end{figure}

\subsection{Challenges in Uncertain String Searching}
We summarize some challenges of searching in uncertain strings.
\begin{itemize}
\item An uncertain string of length $n$ can have multiple characters at each position. As the length of an uncertain string increases, the number of possible worlds grows exponentially. This makes a naive technique that exhaustively enumerates all possible worlds infeasible.

\item Since multiple substrings can be enumerated from the same starting position, care should be taken in substring searching to avoid possible duplication of reported positions. 

\item Enumerating all the possible sequences for arbitrary probability threshold $\tau$ and indexing them requires massive space for large strings. Also note that, for a specific starting position in the string, the probability of occurrence of a substring can change arbitrarily (non-decreasing order) with increasing length, depending on the probability of the concatenated character This makes it difficult to construct index that can support arbitrary probability threshold $\tau$.

\item Correlated uncertainty among the string positions is not uncommon in applications. An index that handles correlation appeals to a wider range of applications. However, handling the correlation can be a bottleneck on space and time.
\end{itemize}

\subsection{Related Work}
Although, searching over clean data has been widely researched, indexing uncertain data is relatively new. Below we briefly mention some of the previous works related to uncertain strings.

\begin{description}
\item[Algorithmic Approach]
Li et al.~\cite{LiBKP14} tackled the substring searching problem where both the query substring and uncertain sequence comes as online query. They proposed a dynamic programming approach to calculate the probability that a substring is contained in the uncertain string. Their algorithm takes linear time and linear space. 

\item[Approximate substring Matching] Given as input a string $p$, a set of strings \{${x_i | 1 \leq i \leq r}$\}, and an edit distance threshold $k$, the substring matching problem is to find all substrings $s$ of $x_i$ such that $d(p,s)\leq k$, where $d(p,s)$ is the edit distance between $p$ and $s$. This problem has been well studied on clean texts (see~\cite{Navarro01} for a survey). Most of the ideas to solve this problem is based on partitioning $p$. Tiangjian et al.~\cite{GeL11} extended this problem for uncertain strings.  Their index can handle strings of arbitrary lengths.

\item[Frequent itemset mining]
Some articles discuss the problem of frequent itemset mining in uncertain databases~\cite{ChuiK08,ChuiKH07,LeungH09,BerneckerKRVZ12}, where an itemset is called frequent if the probability of occurrence of the itemset is above a given threshold. 

\item[Probabilistic Database]
Several works (~\cite{cheng2004efficient,tao2005indexing,singh2007indexing}) have developed indexing techniques for probabilistic databases, based on R-trees and inverted indices, for efficient execution of nearest neighbor queries and probabilistic threshold queries. Dalvi et al.~\cite{dalvi2007efficient} proposed efficient evaluation method for arbitrary complex SQL queries in probabilistic database. Later on efficient index for ranked top-$k$ SQL query answering on a probabilistic database was proposed(~\cite{re2007efficient,li2011unified}). Kanagal et al.~\cite{kanagal2009indexing} developed efficient data structures and indexes for supporting inference and decision support queries over probabilistic databases containing correlation. They use a tree data structure named junction tree to represent the correlations in the probabilistic database over the tuple-existence or attribute-value random variables.

\item[Similarity joins]
A string similarity join finds all similar string pairs between two input string collections. Given two collections of
uncertain strings $R$, $S$, and input $(k, \tau )$, the task is to find string pairs $(r, s)$ between these collections
such that $Pr(ed(R, S)\leq k) > \tau$ i.e., probability of edit distance between $R$ and $S$ being at most $k$ is more than probability threshold $\tau$. There are some works on string joins, e.g., (~\cite{ChaudhuriGK06,GravanoIJKMS01,LiLL08}), involving approximation, data cleaning, and noisy keyword search, which has been discussed in the probabilistic setting(~\cite{JestesLYY10}). Patil et al.~\cite{patil2014similarity} introduced filtering techniques to give upper and (or) lower bound on $Pr(ed(R, S)\leq k)$ and incorporate such techniques into an indexing scheme with reduced filtering overhead.
\end{description} 

\subsection{Our Approach}

Since uncertain string indexing is more complex than deterministic string indexing, a general solution for substring searching is challenging. However efficiency can be achieved by tailoring the data structure based on some key parameters, and use the data structure best suited for the purposed application. We consider the following parameters for our index design.

\textbf{Threshold parameter $\tau_{min}$. }
The task of substring matching in uncertain string is to find all the probable occurrences, where the probable occurrence is determined by a query threshold parameter $\tau$. Although $\tau$ can have any value between $0$ to $1$ at query time, real life applications usually prohibits arbitrary small value of $\tau$. For example, a monitoring system does not consider a sequence of events as a real threat if the associated probability is too low. We consider a threshold parameter $\tau_{min}$, which is a constant known at construction time, such that query $\tau$ does not fall below $\tau_{min}$. Our index can be tailored based on $\tau_{min}$ at construction time to suit specific application needs.
 
\textbf{Query substring length. } 
The query substring searched in the uncertain string can be of arbitrary length ranging from 1 to $n$. However, most often the query substrings are smaller than the indexed string. An example is a sensor system, collecting and indexing big amount of data to facilitate searching for interesting query patterns, which are smaller compared to the data stream. We show that more efficient indexing solution can be achieved based on query substring length.

\textbf{Correlation among string positions. } 
Probabilities at different positions in the uncertain string can possibly contain correlation among them. In this paper we consider character level uncertainly model, where a probability of occurrence of a character at a position can be correlated with occurrence of a character at a different position. We formally define the correlation model and show how correlation is handled in our indexes.

Our approach involves the use of suffix trees, suffix arrays and range maximum query data structure, which to the best of our knowledge, is the first use for uncertain string indexing. Succinct and compressed versions of these data structures are well known to have good practical performance. Previous efforts to index uncertain strings mostly involved dynamic programming and lacked theoretical bound on query time. We also formulate the uncertain string listing problem. Practical motivation for this problem is given in Section~\ref{sec_docretrieval}. As mentioned before, for a specific starting position of an uncertain string, the probability of occurrence of a substring can change arbitrarily with increasing length depending on the probability of the concatenated character. We propose an approximate solution by discretizing the arbitrary probability changes with conjunction of a linking structure in the suffix tree.
  
\subsection{Our Contribution:}
In this paper, we propose indexing solutions for substring searching in a single uncertain string, searching in a uncertain string collection, and approximate index for searching in uncertain strings. More specifically, we make the following contributions:

\begin{enumerate}
\item For the substring searching problem, we propose a linear space solution for indexing a given uncertain string $S$ of length $n$, such that all the occurrences of a deterministic query string $p$ with probability of occurrence greater than a query threshold $\tau$ can be reported. We show that for frequent cases our index achieves optimal query time proportional to the substring length and output size. Our index can be designed to support arbitrary probability threshold $\tau \geq \tau_{min}$, where $\tau_{min}$ is a constant known at index construction time.

\item For the uncertain string listing problem, given a collection of uncertain strings $\D=\{d_1,\dots ,d_D\}$ of total size $n$, we propose a linear space and near optimal time index for retrieving all the uncertain strings that contain a deterministic query string $p$ with probability of occurrence greater than a query threshold $\tau$. Our index supports queries for arbitrary $\tau \geq \tau_{min}$, where $\tau_{min}$ is a constant known at construction time.

\item We propose an index for approximate substring searching, which can answer substring searching queries in uncertain strings for arbitrary $\tau \geq \tau_{min}$ in optimal $O(m+occ)$ time, where $\tau_{min}$ is a constant known at construction time and $\epsilon$ is the bound on desired additive error in the probability of a matched string, i.e. outputs can have probability of occurrence $\geq \tau-\epsilon$.

\end{enumerate}



\subsection{Outline}
The rest of the paper is organized as follows. In section~\ref{sec_motivation} we show some practical motivations for our indexes. In section~\ref{preliminaries} we give a formal definition of the problem, discuss some definitions related to uncertain strings and supporting tools used in our index. In section~\ref{sec_special} we build a linear space index for answering a special form of uncertain strings where each position of the string has only one probabilistic character. In section~\ref{sec_general} we introduce a linear space index to answer substring matching queries in general uncertain strings for variable threshold. Section~\ref{sec_docretrieval} discusses searching in an uncertain string collection. In section~\ref{sec_apm}, we discuss approximate string matching in uncertain strings. In section~\ref{sec_experiments}, we show the experimental evaluation of our indexes. Finally in section~\ref{conclusion}, we conclude the paper with a summary and future work direction.

\section{Motivation}
\label{sec_motivation}
Various domains such as bioinformatics, knowledge discovery for moving object database trajectories, web log analysis, text mining, sensor networks, data integration and activity recognition generates large amount of uncertain data. Below we show some practical motivation for our indexes.

\begin{description}
\item[Biological sequence data]
Sequence data in bioinformatics is often uncertain and probabilistic. For instance, reads in shotgun sequencing are annotated with quality scores for each base. These quality scores can be understood as how certain a sequencing machine is about a base. Probabilities over long strings are also used to represent the distribution of SNPs or InDels (insertions and deletions) in the population of a species. Uncertainty can arise due to a number of factors in the high-throughput sequencing technologies. NC-IUB committee standardized incompletely specified bases in DNA(~\cite{lilley1996nomenclature}) to address this common presence of uncertainty. Analyzing these uncertain sequences is important and more complicated than the traditional string matching problem.  

We show an example uncertain string generated by aligning genomic sequence of Tree of At4g15440 from OrthologID and deterministic substring searching in the sequence. Figure~\ref{genome} illustrates the example.

\begin{figure}[h!]
\begin{center}
	\tiny
    \begin{tabular}{ccccccccccccccc}
    S[1] &S[2]&S[3]&S[4] &S[5]&S[6] &S[7] &S[8]&S[9] &S[10] &S[11]\\
    P 1 &S .7 &F 1 &P 1 &Q .5 &P 1 &A .4 &I .3 &A 1 &S .5 &A 1  \\ 
    	&F .3 &    &    &T .5 &    &F .4 &L .3 &    &T .5 &     \\ 
    	&     &    &    &     &    &P .2 &P .3 &    &     &     \\
    	&     &    &    &     &    &     &T .3 &    &     &     \\
    	
    \end{tabular}
   
\end{center}
 \caption{Example of an uncertain string $S$ generated by aligning genomic sequence of the tree of At4g15440 from OrthologID.}
\label{genome}
\end{figure}

Consider the uncertain string $S$ of Figure~\ref{genome}. A sample query can be $\{p="AT",\tau=0.4\}$, which asks to find all the occurrences of string $AT$ in $S$ having probability of occurrence more than $\tau=.4$. $AT$ can be matched starting at position $7$ and starting at position $9$. Probability of occurrence for starting position $7$ is $0.4\times 0.3=0.12$ and for starting position $9$ is $1\times 0.5=0.5$. Thus position $9$ should be reported to answer this query.

\item[Automatic ECG annotations]
In the Holter monitor application, for example, sensors attached to heart-disease patients send out ECG signals continuously to a
computer through a wireless network(~\cite{dash2009automatic}). For each heartbeat, the annotation software gives a symbol such as N (Normal beat), L (Left bundle branch block beat), and R, etc. However, quite often, the ECG signal of each beat may have ambiguity, and a probability distribution on a few possibilities can be given. A doctor might be interested in locating a pattern such as “NNAV” indicating two normal beats followed by an atrial premature beat and then a premature ventricular contraction, in order to verify a specific diagnosis. The ECG signal sequence forms an uncertain string, which can be indexed to facilitate deterministic substrings searching.

\item[Event Monitoring]
Substring matching over event streams is important in paradigm where continuously arriving events are matched. For example a RFID-based security monitoring system produces stream of events. Unfortunately RFID devices are error prone, associating probability with the gathered events. A sequence of events can represent security threat. The stream of probabilistic events can be modeled with uncertain string and can be indexed so that deterministic substring can be queried to detect security threats.

\end{description}

\section{Preliminaries}
\label{preliminaries}
\subsection{Uncertain String and Deterministic String}
An uncertain string $S=s_1\dots s_n$ over alphabet $\Sigma$ is a sequence of sets $s_i,i=1,\dots ,n$. Every $s_i$ is a set of pairs of the form $(c_j,pr(c_j^i))$, where every $c_j$ is a character in $\Sigma$ and $0\leq pr(c_j^i)\leq 1$ is the probability of occurrence of $c_j$ at position $i$ in the string. Uncertain string length is the total number of positions in the string, which can be less than the total number of characters in the string. Note that, summation of probability for all the characters at each position should be $1$, i.e. $\sum\limits_{j} pr(c_j^i) = 1$. Figure~\ref{genome} shows an example of an uncertain string of length $11$. A deterministic string has only one character at each position with probability $1$. We can exclude the probability information for deterministic strings.

\subsection{Probability of Occurrence of a Substring in an Uncertain String}
Since each character in the uncertain string has an associated probability, a deterministic substring occurs in the uncertain string with a probability. Let $S=s_1\dots s_n$ is an uncertain string and $p$ is a deterministic string. If the length of $p$ is $1$, then probability of occurrence of $p$ at position $i$ of $S$ is the associated probability $pr(p^i)$. Probability of occurrence of a deterministic substring $p=p_1\dots p_k$, starting at a position $i$ in $S$ is defined as the partial product $pr(p_1^i)\times \dots\times pr(p_k^{i+k-1})$. For example in Figure~\ref{genome}, $SFPQ$ has probability of occurrence $0.7\times 1\times 1\times 0.5=0.35$ at position $2$.

\subsection{Correlation Among String Positions. } 
We say that character $c_k$ at position $i$ is correlated with character $c_l$ at position $j$, if the probability of occurrence of $c_k$ at position $i$ is dependent on the probability of occurrence of $c_l$ at position $j$. We use $pr(c_k^i)^+$ to denote the probability of $c_k^i$ when the correlated character is present, and $pr(c_k^i)^-$ to denote the probability of $c_k^i$ when the correlated character is absent. Let $x_g\ldots x_h$ be a the substring generated from an uncertain string. $c_k^i$,$g\leq i\leq h$ is a character within the substring which is correlated with $c_l^j$. Depending on the position $j$, we have $2$ cases:

\begin{description}
\item[Case 1, $g\leq j\leq h$ : ] The correlated probability of $(c_k^i)$ is expressed by $(c_l^j\implies a$ , $\neg c_l^j\implies b)$, i.e. if $c_l^j$ is taken as an occurrence, then $pr(c_k^i)=pr(c_k^i)^+$, otherwise $pr(c_k^i)=pr(c_k^i)^-$. We consider a simple example in Figure~\ref{correlated_string} to illustrate this. In this string, $z^3$ is correlated with $e^1$. For the substring $eqz$, $pr(z^3)=.3$, and for the substring $fqz$, $pr(z^3)=.4$.

\item[Case 1, $j< g$ or $j> h$ : ]  $c_l^j$ is not within the substring. In this case, $pr(c_k^i)$=$pr(c_l^j)$*$pr(c_k^i)^+$+$(1-pr(c_l^j))$*$pr(c_k^i)^+$. In Figure~\ref{correlated_string}, for substring $qz$, $pr(z^3)=.6*.3+.4*.4$. 
\end{description}

\begin{figure}[h!]
\begin{center}
	\small
    \begin{tabular}{|c|c|c|}
   \hline
    S[1] &S[2]&S[3]\\\hline
    e: .6 &q: 1 &z: $e^1\implies .3,\neg e^1\implies .4$\\ 
    f: .4 &    &    \\ 
    	&     &    \\
    	&     &    \\
\hline
    \end{tabular}
   
\end{center}
 \caption{Example of an uncertain string $S$ with correlated characters.}
\label{correlated_string}
\end{figure}


\subsection{Suffix Trees and Generalized Suffix Trees}
The suffix tree~\cite{st1,st2} of a deterministic string $t[1\dots n]$ is a lexicographic arrangement of all these $n$ suffixes in a compact trie structure of $O(n)$ words space, where the $i$-th leftmost leaf represents the $i$-th lexicographically smallest suffix of $t$. For a node $i$  (i.e., node with pre-order rank $i$), $path(i)$ represents the text obtained by concatenating all edge labels on the path from root to node $i$ in a suffix tree.
The locus node $i_p$ of a string $p$ is the node closest to the root such that the $p$ is a prefix of $path(i_P)$.
The suffix range of a string $p$ is given by the maximal range $[sp, ep]$ such that for $sp \leq j \leq ep$, $p$ is a prefix  of (lexicographically) $j$-th suffix of $t$.
Therefore, $i_p$ is the lowest common ancestor of $sp$-th and $ep$-th leaves.
Using suffix tree, the locus node as well as the suffix range of $p$ can be computed in $O(p)$ time, where $p$ denotes the length of $p$. The suffix array $A$ of $t$ is defined to be an array of integers providing the starting positions of suffixes of $S$ in lexicographical order. This means, an entry $A[i]$ contains the position of $i$-th leaf of the suffix tree in $t$.
For a collection of strings $\D=\{d_1,\dots ,d_D\}$, let $t= d_1d_2\dots d_D$ be the text obtained by concatenating all the strings in $\D$. Each string is assumed to end with a special character \$. The suffix tree of $t$ is called the generalized suffix tree (GST) of $\D$.

\section{String Matching in Special Uncertain Strings}
\label{sec_special}
In this section, we construct index for a special form of uncertain string which is extended later. Special uncertain string is an uncertain string where each position has only one probable character with associated non-zero probability of occurrence. Special-uncertain string is defined more formally below.

\BDE [Special uncertain string]
A special uncertain string $X=x_1\dots x_n$ over alphabet $\Sigma$ is a sequence of pairs. Every $x_i$ is a pair of the form $(c_i,pr(c_i^i))$, where every $c_i$ is a character in $\Sigma$ and $0< pr(c_i^i)\leq 1$ is the probability of occurrence of $c_i$ at position $i$ in the string.
\EDE

Before we present an efficient index, we discuss a naive solution similar to deterministic substring searching.

\subsection{Simple Index} 
\begin{figure}
    \begin{center}
      \includegraphics[width=0.3\textwidth]{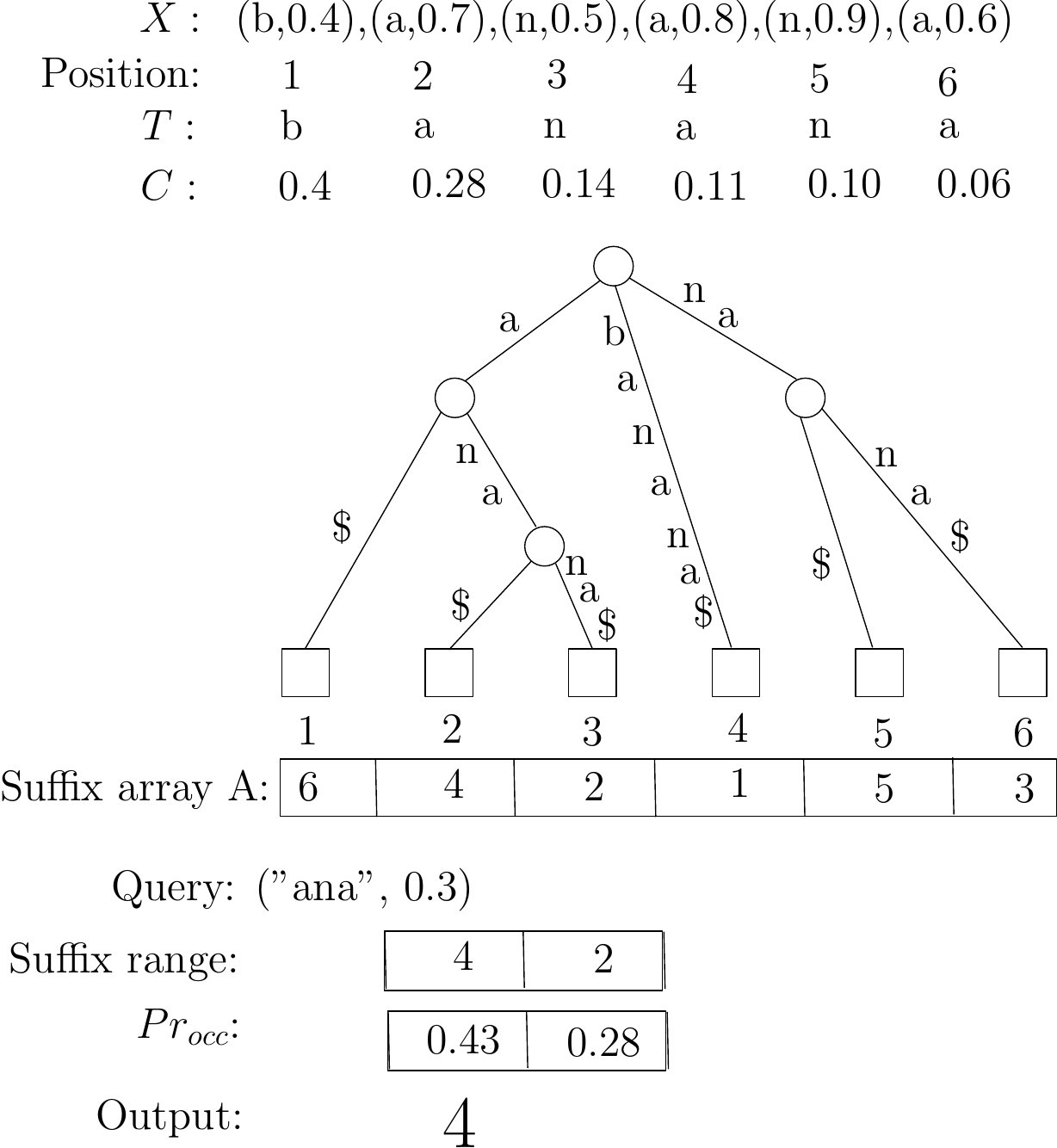}
    \end{center}
    \caption{Simple index for special uncertain strings.}
 \label{fig:simpleindex}
  \end{figure}

Given a special uncertain string $X=x_1\dots x_n$, construct the deterministic string $t=c_1\dots c_n$ where $c_i$ is the character in $x_i$. We build a suffix tree over $t$. We build a suffix array $A$ which maps each leaf of the suffix tree to its original position in $t$. We also build a successive multiplicative probability array $C$, where $C[j]=\prod_{i=1}^j Pr(c_i^i)$, for $j=1,\dots ,n$. For a substring $x_i\dots x_{i+j}$, probability of occurrence can be easily computed by $C[i+j]/C[i-1]$.
Given an input ($p,\tau$), we traverse the suffix tree for $p$ and find the locus node and suffix range of $p$ in $O(m)$ time, where $m$ is the length of $p$. Let the suffix range be $[sp, ep]$. According to the property of suffix tree, each leaf within the range $[sp, ep]$ contains an occurrence of $p$ in $t$. Original positions of the occurrence in $t$ can be found using suffix array, i.e., $A[sp],\dots ,A[ep]$. However, each of these occurrence has an associated probability. We traverse each of the occurrence in the range $A[sp],\dots ,A[ep]$. For an occurrence $A[i]$, we find the probability of occurrence by $C[A[i]+m-1]/C[A[i]-1]$. If the probability of occurrence is greater than $\tau$, we report the position $A[i]$ as an output. Figure~\ref{fig:simpleindex} illustrates this approach.

\textbf{Handling Correlation:}
Correlation is handled when we check for the probability of occurrence. If $A[i]$ is a possible occurrence, then we need to consider any existing character within the substring $x_i\dots x_{i+m-1}$, that is correlated with another character. Let $c_k$ is a character at position $j$ within $x_i\dots x_{i+m-1}$, which is correlated with character $c_l^{j'}$, i.e. if $c_l^{j'}$ is included in the substring, then $pr(c_k^j)$=$pr(c_k^j)^+$, or else $pr(c_k^j)$=$pr(c_k^j)^-$. To find the correct probability of $(c_k^j)$, if $j'$ we check the $j'$-th position ($j'$ depth character on the root to locus path in the suffix tree) of the substring. If the $j'$-th character is $c_l$, then $C[A[i]+m-1]/C[A[i]-1]$ is the correct probability of occurrence for $x_i\dots x_{i+m}$. Otherwise, $C[A[i]+m-1]/C[A[i]-1]$ contains the incorrect probability of $c_k^j$. Dividing $C[A[i]+m-1]/C[A[i]-1]$ by $pr(c_k^j)^+$ and multiplying by $pr(c_k^j)^-$ gives the correct probability of occurrence in this case. If $c_l$ falls before or after the substring $x_i\dots x_{i+m-1}$, $pr(c_k^j)$=$pr(c_l^{j'})$*$pr(c_k^j)^+$+$(1-pr(c_l^{j'}))$*$pr(c_k^j)^-$. Dividing $C[A[i]+m-1]/C[A[i]-1]$ by $pr(c_k^j)^+$ and multiplying by $pr(c_k^j)$ gives the correct probability of occurrence. Note that, we can identify and group all the characters with existing correlation, and search in the suffix tree in one scan for improved efficiency.

The main drawback in this approach is the query time. Within the suffix range $[sp, ep]$, possibly very few number of positions can qualify as output because of $\tau$. So spending time on each element of the range $[sp, ep]$ is not justifiable. 

\subsection{Efficient Index:}
Bottleneck of the simple index comes from traversing each element within the suffix range. For the efficient index, we iteratively retrieve the element with maximum probability of occurrence in the range in constant time. Whenever the next maximum  probability of occurrence falls below $\tau$, we conclude our search. We use range maximum query (RMQ) data structure for our index which is briefly explained below. 

\textbf{Range Maximum Query}

Let $B$ be an array of integers of length $n$, a range maximum query($RMQ$) asks for the position of the maximum value between two specified array indices $[i, j]$. i.e., the $RMQ$ should return an index $k$ such that $i \leq k \leq j$ and $B[k] \geq B[x]$ for all $i \leq x \leq j$. We use the result captured in following lemma for our purpose.

 \BL
\label{RMQ}~\cite{FHS07,FHS08}
By maintaining a $2n+o(n)$ bits structure, range maximum query(RMQ) can be answered in $O(1)$ time (without accessing the array).
\EL

Every leaf of the suffix tree denotes a suffix position in the original text and a root to leaf path represents the suffix. For uncertain string, every character in this root to leaf path has an associated probability which is not stored in the suffix tree.
Let $y_j^i$, for $j=1,\dots, n$ denote a deterministic substring which is the $i$-length prefix of the $j$-th suffix,i.e. the substring on the root to $i$-th leaf path. Let $Y^i$ is the set of $y_j^i$, for $j=1,\dots, n$.

For $i=1,\dots, n$, we define $C_i$ as the successive multiplicative probability array for the substrings of $Y^i$. $j$-th element of $C_i$ is the successive multiplicative probability of the $i$-length prefix of the $j$-th suffix. More formally $C_i[j]=\prod_{k=A[j]}^{A[j]+i-1} Pr(c_k^k)=C[A[j]+i-1]/C[A[j]-1] (1\leq j\leq n$). For each $C_i (i=1,\dots ,\log n$) we use range maximum query data structure $RMQ_i$ of $n$ bits over $C_i$ and discard the original array $C_i$. Note that, $RMQ$ data structure can be built over an integer array. We convert $C_i$ into an integer array by multiplying each element by a sufficiently large number and then build the $RMQ_i$ structure over it. We obtain $\log n$ number of such $RMQ$ data structures resulting in total space of $O(n\log n)$ bits or $O(n)$ words. We also store the global successive multiplicative probability array $C$, where $C[j]=\prod_{i=1}^j Pr(c_i^i)$. Given a query $(p,\tau)$, Our idea is to use $RMQ_i$ for iteratively retrieving maximum probability of occurrence elements in constant time each and validate using $C$. To maintain linear space, we can support query substring length of $m=0,\dots ,\log n$ in this approach. Algorithm~\ref{alg_special1} illustrates the index construction phase for short substrings.

\begin{center}
\textbf{Query Answering}
\end{center}
\begin{description}
\item[Short substrings ($m\leq \log n$) : ]
Given an input $(p, \tau)$, we first retrieve the suffix range $[l,r]$ in $O(m)$ time using suffix tree, where $m$ is the length of $p$. We can find the maximum probability occurrence of $p$ in $O(1)$ time by executing query $RMQ_m(l,r)$. Let $max$ be the position of maximum probability occurrence and $max'=A[max]$ be the the original position in $t$. We can find the corresponding probability of occurrence by $C[max'+i-1]/C[max'-1]$. If the probability is less that $\tau$, we conclude our search. If it is greater than $\tau$, we report $max'$ as an output. For finding rest of the outputs, we recursively search in the ranges $[l,max-1]$ and $[max+1,r]$. Since each call to $RMQ_m(l,r)$ takes constant time, validating the probability of occurrence takes constant time, we spend $O(1)$ time for each output. Total query time is optimal $O(m+occ)$. Algorithm~\ref{alg_special2} illustrates the query answering for short substrings. Note that, correlation is handled in similar way as described for the naive index, and we omit the details here.

\begin{algorithm}
\SetKwInOut{Input}{input}
\SetKwInOut{Output}{output}
 \Input{A special uncertain string $X$}
 \Output{suffix tree, suffix array $A$, successive multiplicative probability array $C$, $RMQ_i$ ($i=1,\dots,\log n$)}
 Build deterministic string $t$ from $X$\\
 Build suffix tree over $t$\\
 Build suffix array $A$ over $t$\\
 \tcp{Building successive multiplicative probability array}
 $C[1]=Pr(c_1^1)$\\
 \For{$i=2;i\leq n;i++$}{
 	$C[i]=C[i-1]\times Pr(c_i^i)$ 
 }
 \tcp{Building $C_i$ array for $i=1,\dots ,\log n$}
 \For{$i=1;i\leq \log n;i++$}{
 	\For{$j=1;j\leq n;j++$}{
 		$C_i[j]=C[A[j]+i-1]/C[A[j]-1]$\\
 		 \tcp{Handling correlated characters}
 		\For{all character $c_a^k$ in $t[A[j]\dots t[A[j]+i-1]$ that are correlated with another character $c_b^l$}{
 		\textbf{if} ($A[j]\leq l\leq[A[j]+i-1]$ and $c_b^l$ is not within $t[A[j]\dots t[A[j]+i-1]$)
 		$C_i[j]=C_i[j]/Pr(c_a^k)^+*Pr(c_a^k)^-$\\
 		\textbf{else}\\
 		$pr(c_a^k)$=$pr(c_b^l)$*$pr(c_a^k)^+$+$(1-pr(c_b^l))$*$pr(c_a^k)^-$\\
 		$C_i[j]=C_i[j]/Pr(c_a^k)^+*Pr(c_a^k)$\\
 		\textbf{end}
 		}
 	}
 	Build $RMQ_i$ over $C_i$\\
 }

 \caption{Algorithm Special-Short-Substring-Index-Construction}
 \label{alg_special1}
\end{algorithm}

\begin{algorithm}
\SetKwInOut{Input}{input}
\SetKwInOut{Output}{output}
 \Input{Query substring $p$, probability threshold $\tau)$}
 \Output{Occurrence positions of $p$ in $X$ with probability of occurrence greater than $\tau$}
 $m=length(p)$\\
 call RecursiveRmq($m,1,n$)\\

 \begin{algorithmic}[]
\Function{RecursiveRmq}{$i,l,r$}\Comment{Recursive RMQ method}
	\State $max=RMQ_m(l,r)$
	\State $max'=A[max]$
	\State \If{$C[max'+i-1]/C[max'-1]>\tau$}{
	  \State Output $max'$
      \State Call $RecursiveRmq(m,l,max-1)$
      \State Call $RecursiveRmq(m,max+1,r)$
      }
    \EndFunction
    
\end{algorithmic}

 \caption{Algorithm Special-Short-Substring-Query-Answering}
 \label{alg_special2}
\end{algorithm}

\item[Long substrings ($m> \log n$) : ]
We use a blocking scheme for answering long query substrings ($m>\log n$). Since exhaustively enumerating all possible substrings and storing the probabilities for each of them is infeasible, we only store selective probability values at construction time and compute the others at query time. We partition the entire suffix range of suffix array into different size blocks. More formally, for $i=\log n,\dots ,n$, we divide the suffix range $[1,n]$ of suffix array $A[1,n]$ into $O(n/i)$ number of blocks each of size $i$. Let $B_i$ be the set of length $i$ blocks, i.e. $B_i$=\{$[A[1]...A[i]],[A[i+1]...A[2i]],\dots [A[n-i+1]...A[n]]$\} and let $B$=\{$B_{\log n},\dots ,B_n$\}. For a suffix starting at $A[j]$ and for $B_i$, we only consider the length $i$ prefix of that suffix, i.e. $A[j\dots j+i]$. The idea is to store only the maximum probability value per block. For $B_i,i=\log n,\dots ,n$, we define a probability array $PB_i$ containing $n/i$ elements. $PB_i[j]$ is the maximum probability of occurrence of all the substrings of length $i$ belonging to the $j$-th block of $B_i$. We build a range maximum query structure $RMQ_i$ for $PB_i$. $RMQ_i$ takes $O(n/i)$ bits, total space is bounded by $\sum\limits_{i} O(n/i)=O(n\log n)$ bits or $O(n)$ words.

For a query $(p, \tau)$, we first retrieve the suffix range $[l,r]$. This suffix range can spread over multiple blocks of $B_m$. We use $RMQ_m$ to proceed to next step. Note that $RMQ_m$ consists of $N/m$ bits, corresponding to the $N/m$ blocks of $B_m$ in order. Our query proceeds by executing range maximum query in $RMQ_m(l,r)$, which will give us the index of the maximum probability element of string length $m$ in that suffix range. Let the maximum probability element position in $RMQ_m$ is $max$ and the block containing this element is $B_{max}$. Using $C$ array, we can find out if the probability of occurrence is greater than $\tau$. Note that, we only stored one maximum element from each block. If the maximum probability found is greater than $\tau$, we check all the other elements in that block in $O(m)$ time. In the next step, we recursively query $RMQ_m(l,max-1)$ and $RMQ(max+1,r)$ to find out subsequent blocks. Whenever $RMQ$ query for a range returns an element having probability less than $\tau$, we stop the recursion in that range. Number of blocks visited during query answering is equal to the number of outputs and inside each of those block we check $m$ elements, obtaining total query time of $O(m\times occ)$.
\end{description}
%

In practical applications, query substrings are rarely longer than $\log n$ length. Our index achieves optimal query time for substrings of length less than $\log n$. We show in the experimental section that on average our index achieves efficient query time proportional to substring length and number of outputs reported.

\section{Substring Matching in General Uncertain String}
\label{sec_general}
In this section we construct index for general uncertain string based on the index of special uncertain string. The idea is to convert a general uncertain string into a special uncertain string, build the data structure similar to the previous section and carefully eliminate the duplicate answers. Below we show the steps of our solution in details.

\subsection{Transforming General Uncertain String}
We employ the idea of Amihood et al~\cite{AmirCIKZ08} to transform general uncertain string into a special uncertain string. \textbf{Maximal factor} of an uncertain string is defined as follows.

 \BDE
A \textbf{Maximal factor} of a uncertain string $S$ starting at location $i$ with respect to a fixed probability threshold $\tau_c$ is a string of maximal length that when aligned to location $i$ has probability of occurrence at least $\tau_c$.
 \EDE
 
For example in figure~\ref{genome}, maximal factors of the uncertain string $S$ at location $5$ with respect to probability threshold $0.15$ are "QPA", "QPF", "TPA", "TPF". 
 
 An uncertain string $S$ can be transformed to a special uncertain string by concatenating all the maximal factors of $S$ in order. Suffix tree built over the concatenated maximal factors can answer substring searching query for a fixed probability threshold $\tau_c$. But this method produces a special uncertain string of $\Omega(n^2)$ length, which is practically infeasible. To reduce the special uncertain string length, Amihood et al.~\cite{AmirCIKZ08} employs further transformation to obtain a set of extended maximal factors. Total length of the extended maximal factors is bounded by $O((\frac{1}{\tau_c})^2n)$. 

\BL
\label{Extended}
Given a fixed probability threshold value $\tau_c (0<\tau_c\leq 1)$, an uncertain string $S$ can be transformed into a special uncertain string $X$ of length $O((\frac{1}{\tau_c})^2n)$ such that any deterministic substring $p$ of $S$ having probability of occurrence greater than $\tau_c$ is also a substring of $X$.
\EL 
 
Simple suffix tree structure for answering query does not work for the concatenated extended maximal factors. A special form of suffix tree, namely property suffix tree is introduced by Amihood et al.~\cite{AmirCIKZ08}. Also substring searching in this method works only on a fixed probability threshold $\tau_c$. A naive way to support arbitrary probability threshold is to construct special uncertain string and property suffix tree index for all possible value of $\tau_c$, which is practically infeasible due to space usage.

We use the technique of lemma~\ref{Extended} to transform a given general uncertain string to a special uncertain string of length $O((\frac{1}{\tau_{min}})^2n)$ based on a probability threshold $\tau_{min}$ known at construction time, and employ a different indexing scheme over it. Let $X$ be the transformed special uncertain string. See Figure~\ref{queryappendix} for an example of the transformation. Following section elaborates the subsequent steps of the index construction.

\subsection{Index Construction on the Transformed Uncertain String}
Our index construction is similar to the index of section~\ref{sec_special}. We need some additional components to eliminate duplication and position transformation.

Let $N=|X|$ be the length of the special uncertain string $X$. Note that $N=O((\frac{1}{\tau_{min}})^2n)=O(n)$, since $\tau_{min}$ is a constant known in construction time. For transforming the positions of $X$ into the original position in $S$, we store an array $Pos$ of size $N$, where $Pos[i]$=position of the $i$-th character of $X$ in the original string $S$. We construct the deterministic string $t=c_1\dots c_N$ where $c_i$ is the character in $X_i$. We build a suffix tree over $t$. We build a suffix array $A$ which maps each leaf of the suffix tree to its position in $t$. We also build a successive multiplicative probability array $C$, where $C[j]=\prod_{i=1}^j Pr(c_i^i)$, for $1\leq j\leq N$. For a substring of length $j$ starting at position $i$, probability of occurrence of the substring in $X$ can be easily computed by $C[i+j-1]/C[i-1]$. For $i=1,\dots, n$, we define $C_i$ as the successive multiplicative probability array for substring length $i$ i.e. $C_i[j]=\prod_{k=A[j]}^{A[j]+i-1} Pr(c_k^k)=C[A[j]+i-1]/C[A[j]-1]$ ($1\leq j\leq n$). Figure~\ref{queryappendix} shows $Pos$ array and $C$ array after transformation of an uncertain string. Below we explain how duplicates may arise in outputs and how to eliminate them.

Possible duplicate positions in the output arises because of the general to special uncertain string transformation. Note that, distinct positions in $X$ can correspond to the same position in the original uncertain string $S$, resulting in same position possibly reported multiple times. A key observation here is that for two different substrings of length $m$, if the locus nodes are different than the corresponding suffix ranges are disjoint. These disjoint suffix ranges collectively cover all the leaves of the suffix tree. For each such disjoint ranges, we need to store probability values for only the unique positions of $S$. Without loss of generality we store the value for leftmost unique position in each range. 

For any node $u$ in the suffix tree, $depth(u)$ is the length of the concatenated edge labels from root to $u$. We define by $L_i$ as the set of nodes $u_i^j$ such that $depth(u_i^j)\geq i$ and $depth(parent(u_i^j))\leq i$. For $L_i={u_i^1,\dots ,u_i^k}$, we have a set of disjoint suffix ranges ${[sp_i^1,ep_i^1],\dots ,[sp_i^k,ep_i^k]}$. A suffix range $[sp_i^j,ep_i^j]$ can contain duplicate positions of $S$. Using the $Pos$ array we can find the unique positions for each range and store only the values corresponding to the unique positions in $C_i$.

We use range maximum query data structure $RMQ_i$ of $n$ bits over $C_i$ and discard the original array $C_i$. Note that, $RMQ$ data structure can be built over an integer array. We convert $C_i$ into an integer array by multiplying each element by a sufficiently large number and then build the $RMQ_i$ structure over it. We obtain $\log n$ number of such $RMQ$ data structures resulting in total space of $O(n\log n)$ bits or $O(n)$ words. For long substrings ($m>\log n$), we use the blocking data structure similar to section~\ref{sec_special}. Algorithm~\ref{alg_generalCon} illustrates the index construction phase for short substrings.

\subsection{Query Answering}
Query answering procedure is almost similar to the query answering procedure of section~\ref{sec_special}. Only difference being the transformation of position which is done using the $Pos$ array. Algorithm~\ref{alg_generalQuery} illustrates the query answering phase for short query substrings. See Figure~\ref{queryappendix} for a query answering example.

\subsection{Space Complexity}
For analyzing the space complexity, we consider each component of our index. Length of the special uncertain string $X$ and deterministic string $t$ are $O(n)$, where $n$ is the number of positions in $S$. Suffix tree and suffix tree each takes linear space. We store a successive probability array of size $O(n)$. We build probability array $C_i$ for $i=1,\dots ,\log n$, where each $C_i$ takes of $O(n)$. However we build $RMQ_i$ of $n$ bits over $C_i$ and discard the original array $C_i$. We obtain $\log n$ number of such $RMQ$ data structures resulting in total space of $O(n\log n)$ bits or $O(n)$ words. For the blocking scheme, we build  $RMQ_i$ data structure for $i=\log n,\dots ,n$. $RMQ_i$ takes $n/i$ bits, total space is $\sum\limits_{i} n/i=O(n\log n)$ bits or $O(n)$ words. Since each component of our index takes linear space, total space taken by our index is $O(n)$ words.

\subsection{Proof of Correctness}
In this section we discuss the correctness of our indexing scheme. 
\textbf{Substring conservation property of the transformation : }
At first we show that any substring of $S$ with probability of occurrence greater than query threshold $\tau$ can be found in $t$ as well. 
According to lemma~\ref{Extended}, a substring having probability of occurrence greater than $\tau_{min}$ in $S$ is also a substring of the transformed special uncertain string $X$. Since query threshold value $\tau$ is greater than $\tau_{min}$, and entire character string of $X$ is same as the deterministic string $t$, a substring having probability of occurrence greater than query threshold $\tau$ in $S$ will be present in the deterministic string $t$.

\textbf{Algorithm~\ref{alg_generalQuery} outputs the complete set of occurrences : } 
For contradiction, we assume that an occurrence position $z$ of substring $p$ in $S$ having probability of occurrence greater than $\tau$ is not included in the output. From the aforementioned property, $p$ is a substring of $t$. According to the property of suffix tree, $z$ must be present in the suffix range $[sp,ep]$ of $p$. Using $RMQ$ structure, we report all the occurrence in $[sp,ep]$ in their decreasing order of probability of occurrence value in $S$ and stop when the probability of occurrence falls below $\tau$, which ensures inclusion of $z$.

\textbf{Algorithm~\ref{alg_generalQuery} does not output any incorrect occurrence : } 
An output $z$ can be incorrect occurrence if it is not present in uncertain string $S$ or its probability of occurrence is less than $\tau$. We query only the occurrences in the suffix range $[sp,ep]$ of $p$, according to the property of suffix tree all of which are valid occurrences. We also validate the probability of occurrence for each of them using the successive multiplicative probability array $C$.


\section{String Listing from Uncertain String Collection}
\label{sec_docretrieval}
In this section we propose an indexing solution for problem~\ref{def:retrieval}. We are given a collection of $D$ uncertain strings $\D=\{d_1,\dots ,d_D\}$ of $n$ positions in total. Let $i$ denotes the string identifier of string $d_i$. For a query $(p,\tau)$, we have to report all the uncertain string identifiers $j$ such that $d_j$ contains $p$ with probability of occurrence more than $\tau$. In other words, we want to list the strings from a collection of a string, that are relevant to a deterministic query string based on probability parameter.

\textbf{Relevance metric : }
For a deterministic string $t$ and an uncertain string $S$, we define a relevance metric, $Rel(S,t)$. If $t$ does not  have any occurrence in $S$, then $Rel(S,t)$=$0$. If $s$ has only one occurrence of $t$, then $Rel(S,t)$ is the probability of occurrence of $t$ in $S$. If $s$ contains multiple occurrences of $t$, then $Rel(S,t)$ is a function of the probability of occurrences of $t$ in $S$. Depending on the application, various functions can be chosen as the appropriate relevance metric. A common relevance metric is the maximum probability of occurrence, which we denote by $Rel(S,t)_{max}$. The $OR$ value of the probability of occurrences is another useful relevance metric. More formally, if a deterministic string $t$ has nonzero probable occurrences at positions $i_1,\dots ,i_k$ of an uncertain string $S$, then we define the relevance metric of $t$ in $S$ as $Rel(S,t)_{OR}$ = $\sum\limits_{j=i_1}^{i_k} pr(t_j) - \prod_{j=i_1}^{i_k} pr(t_j) $, where $pr(t_j)$ is the probability of occurrence of $t$ in $S$ at position $j$. Figure~\ref{figstringlisting2} shows an example. 

\begin{figure}[h!]
\begin{center}
\normalsize Uncertain string $S$:\\
       \begin{tabular}{c}

    \end{tabular}
    \begin{tabular}{|c|c|c|c|c|c|}
    \hline
    $S[1]$ &$S[2]$ &$S[3]$ &$S[4]$ &$S[5]$ &$S[6]$\\
    A .4     &B .3     &A .5     &A .6     &B .5     &A .4\\ 
    B .3     &L .3     &F .5     &B .4     &F .3     &C .3\\ 
    F .3     &F .3     &         &         &J .2     &E .2\\
    	     &J .1     &         &         &         &F .1\\

    \hline
    \end{tabular}
   
       \begin{tabular}{c}
\\
	$Rel(S,"BFA")_{max}$=$.09$\\
    $Rel(S,"BFA")_{OR}$=$(.06+.09+.048)-(.06*.09*.048)$=$.19786$
    \end{tabular}
\end{center}
 \caption{Relevance metric for string listing.}
\label{figstringlisting2}
\end{figure}

\textbf{Practical motivation : }
Uncertain string listing finds numerous practical motivation. Consider searching for a virus pattern in a collection of text files with fuzzy information. The objective is to quarantine the files that contain the virus pattern. This problem can be modeled as a uncertain string listing problem, where the collection of text files is the uncertain string collection $D$, the virus pattern is the query pattern $P$, and $\tau$ is the confidence of matching. Similarly, searching for a gene pattern in genomic sequences of different 
species can be solved using uncertain string listing data structure.

\textbf{The index : }
As explained before, a naive search on each of the string will result in $O(\sum \limits_{i}$search time on $d_i)$ which can be much larger than the actual number of strings containing the string. Objective of our index is to spend only one search time and time proportional to the number of output strings. We construct a generalized suffix tree so that we have to search for the string only once. We concatenate $d_1,\dots ,d_D$ by a special symbol which is not contained in any of the document and obtain a concatenated general uncertain string $S=d_1\$\dots \$d_D$. Next we use the transformation method described in previous section to obtain deterministic string $t$, construct suffix tree and suffix array for $t$. According to the property of suffix tree, the leaves under the locus of a query substring $t$ contains all the occurrence positions of $t$. However, these leaves can possibly contain duplicate positions and multiple occurrence of the same document. In the query answering phase, duplicate outputs can arise because of the following two reasons:
\begin{enumerate}
\item Distinct positions in $t$ can correspond to the same position in the original uncertain string $S$

\item Distinct positions in $S$ can correspond to the same string identifier $d_j$ which should be reported only once

\end{enumerate}

Duplicate elimination is important to keep the query time proportional to the number of output strings. At first we construct the successive multiplicative probability array $C_i$ similar to the substring searching index, then show how to incorporate $Rel(S,t)$ value for the multiple occurrences cases in the same document and duplicate elimination. 

Let $y_j^i$, for $j=1,\dots, n$ denote a deterministic substring which is the $i$-length prefix of the $j$-th suffix,i.e. the substring on the root to $i$-th leaf path. Note that, multiple $y_j^i$ can belong to the same locus node in the suffix tree. Let $Y^i$ is the set of $y_j^i$, for $j=1,\dots, n$. The $i$-depth locus nodes in the suffix tree constitutes disjoint partitions in $Y^i$. For $i=1,\dots, n$, we define $C_i$ as the successive multiplicative probability array for the substrings of $Y^i$. $j$-th element of $C_i$ is the successive multiplicative probability of the $i$-length prefix of the $j$-th suffix. More formally $C_i[j]=\prod_{k=A[j]}^{A[j]+i-1} Pr(c_k^k)=C[A[j]+i-1]/C[A[j]-1] (1\leq j\leq n$). 

The $i$-depth locus nodes in the suffix tree constitutes disjoint partitions in $C_i$. Let $u$ be a $i$-depth locus node having suffix range $[j\dots k]$ and root to $u$ substring $t$. Then the partition $C_i[j\dots k]$ belongs to $u$. For this partitions, we store only one occurrence of a string $d_j$ with the relevance metric value $Rel(S,t)$, and discard the other occurrences of $d_j$ in that range. We build $RMQ$ structure similar to section~\ref{sec_general}.

\textbf{Query Answering}
We explain the query answering for short substrings. Blocking scheme described in previous section can be used for longer query substrings. Given an input $(p, \tau)$, we first retrieve the suffix range $[l,r]$ in $O(m)$ time using suffix tree, where $m$ is the length of $p$. We can find the maximum relevant occurrence of $p$ in $O(1)$ time by executing query $RMQ_m(l,r)$. Let $max$ be the position of maximum relevant occurrence and $max'=A[max]$ be the the original position in $t$. For relevance metric $Rel(S,t)_{max}$, we can find the corresponding probability of occurrence by $C[max'+i-1]/C[max'-1]$. In case of the other complex relevance metric, all the occurrences need to be considered to retrieve the actual value of $Rel(S,t)$. If the relevance metric is less that $\tau$, we conclude our search. If it is greater than $\tau$, we report $max'$ as an output. For finding rest of the outputs, we recursively search in the ranges $[l,max-1]$ and $[max+1,r]$. Each call to $RMQ_m(l,r)$ takes constant time. For simpler relevance metrics (such as $Rel(S,t)_{max}$), validating the relevance metric takes constant time. Total query time is optimal $O(m+occ)$. However, for more complex relevance metric, all the occurrences of $t$ might need to be considered, query time will be proportionate to the total number of occurrences.

%

\section{Approximate Substring Searching}
\label{sec_apm}
In this section we introduce an index for approximate substring matching in an uncertain string. As discussed previously, several challenges of uncertain string matching makes it harder to achieve optimal theoretical bound with linear space. We have proposed index for exact matching which performs near optimally in practical scenarios, but achieves theoretical optimal bound only for shorter query strings. To achieve optimal theoretical bounds for any query, we propose an approximate string matching solution. Our approximate string matching data structure answers queries with an additive error $\epsilon$, i.e. outputs can have probability of occurrence $\geq \tau - \epsilon$. 

At first we begin by transforming the uncertain string $S$ into a special uncertain string $X$ of length $N=O((\frac{1}{\tau_{min}})^2n)$ using the technique of lemma~\ref{Extended} with respect to a probability threshold value $\tau_{min}$. We obtain a deterministic string $t$ from $X$ by concatenating the characters of $X$. We build a suffix tree for $t$. Note that, each leaf in the suffix tree has an associated probability of occurrence $\geq \tau_{min}$ for the corresponding suffix. Given a query $p$, substring matching query for threshold $\tau_{min}$ can now be answered by simply scanning the leafs in subtree of locus node $i_p$. We first describe the framework (based on Hon et. al.~\cite{hon2009space}) which supports a specific probability threshold $\tau$ and then extend it for arbitrary $\tau \geq \tau_{min}$.

We begin by marking nodes in the suffix tree with positional information by associating $Pos_{id} \in [1,n]$. Here,  $Pos_{id}$ indicates the starting position in the original string $S$. A leaf node $l$ is marked with a $Pos_{id} = d$ if the suffix represented by $l$ begins at position $d$ in $S$. An internal node $u$ is marked with $d$ if it is the lowest common ancestor of two leaves marked with $d$. Notice that a node can be marked with multiple position ids. For each node $u$ and each of its marked position id $d$, define a link to be a triplet ($origin,target,Pos_{id}$), where $origin = u$, target is the lowest proper ancestor of $u$ marked with $d$, and $Pos_{id}=d$. Two crucial properties of these links are listed below.

\begin{itemize}

\item Given a substring $p$, for each position $d$ in $S$ where $p$ matches with probability $\geq \tau_{min}$, there is a unique link whose origin is in the subtree of $i_p$ and whose target is a proper ancestor of $i_p$, $i_p$ being the locus node of substring $p$.

\item The total number of links is bounded by $O(N)$.

\end{itemize}

Thus, substring matching query with probability threshold $\tau_{min}$ can now be answered by identifying/reporting the links that originate in the subtree of $i_p$ and are targeted towards some ancestor of it. By referring to each node using its pre-order rank, we are interested in links that are stabbed by locus node $i_p$. Such queries can be answered in $O(m+occ)$, where $|p|=m$ and $occ$ is the number of answers to be reported (Please refer to~\cite{hon2009space} for more details). 

As a first step towards answering queries for arbitrary $\tau \geq \tau_{min}$, we associate probability information along with each link. Thus each link is now a quadruple ($origin,target,Pos_{id}, prob$) where first three parameters remain same as described earlier and $prob$ is the probability of $prefix(u)$ matching uncertain string $S$ at position $Pos_{id} = d$. It is evident that for substring $p$ and arbitrary $\tau \geq \tau_{min}$, a link stabbed by locus node $i_p$ with $prob \geq \tau$ corresponds to an occurrence of $p$ in $S$ at position $d$ with probability $\geq \tau$. However, a link stabbed by $i_p$ with $prob < \tau$ can still produce an outcome since $prefix(i_P)$ contains additional characters not included in $p$, which may be responsible for matching probability to drop below $\tau$. Even though we are interested only in approximate matching this observation leads up the next step towards the solution. We partition each link ($origin = u,target=v,Pos_{id}=d, prob$) into multiple links ($or_1= u,tr_1,d, prob_1$), ($or_2=tr_1,tr_2,d, prob_2$), \dots , ($or_k= tr_{k-1},tr_k=v,d, prob_k$) such that $prob_j - prob_{j-1} \leq \epsilon$ for $2 \leq j \leq k$. Here $or_2, \dots , or_k$ may not refer to the actual node in the suffix tree, rather it can be considered as a dummy node inserted in-between an edge in suffix tree. In essence, we move along the path from node $u = or_1$ towards its ancestors one character at a time till the probability difference is bounded by $\epsilon$ i.e., till we reach node $tr_1$. The process then repeats with $tr_1$ as the origin node and so on till we reach the node $v$. It can be see that the total number of links can now be bounded by $O(N/\epsilon)$. In order to answer a substring matching query with threshold $\tau \geq \tau_{min}$, we need to retrieve all the links stabbed by $i_p$ with $prob \geq \tau$. Occurrence of substring $p$ in $S$ corresponding to each such link is then guaranteed  to have its matching probability at-least $\tau-\epsilon$ due to the way links are generated (for any link with ($u,v$) as origin and target probability of $prefix(v)$ matching in $S$ can be more than that of $prefix(v)$ only by $\epsilon$ at the most).

\section{Experimental Evaluation}
\label{sec_experiments}

 \begin{figure*}
   \label{fig:experiment1}
  \pgfplotsset{width=4.3cm}
   \begin{tikzpicture}
 \begin{axis}[legend pos=north west,
 legend style={draw=none, font=\tiny},
 title=(a)String size vs time,
 xlabel style={ font=\tiny},
 ylabel style={ font=\tiny},
 xlabel={$n\times 1000$},
 ylabel={Query Time(ms)$\times 1000$},
 ymin=.2, ymax=2.8,
 xmin=20, xmax=320,
 legend entries={$\theta=.1$,$\theta=.2$,$\theta=.3$,$\theta=.4$},
 ]
 \addplot+[sharp plot] coordinates
 {(25,.335714) (50,.374308) (100,.458582) (150,.569882) (200,.699356) (250,.702752) (300,.803371)};
 \addplot+[sharp plot] coordinates
 {(25,.385692) (50,.494395) (100,.478703) (150,.639857) (200,.739613) (250,.802163) (300,.853752)};
 \addplot+[sharp plot] coordinates
 {(25,.420305) (50,.554604) (100,.498645) (150,.799436) (200,.819623) (250,.812354) (300,.933775)};
 \addplot+[sharp plot] coordinates
 {(25,.465758) (50,.574108) (100,.604230) (150,.805612) (200,.885623) (250,.962143) (300,1.03156)};
 \end{axis}
 \end{tikzpicture}
 \hskip -5pt
  \begin{tikzpicture}
 \begin{axis}[
 xticklabel style={
 /pgf/number format/precision=2,
 /pgf/number format/fixed,
 /pgf/number format/fixed zerofill,
 },
 legend pos=north east,
 legend style={draw=none, font=\tiny},
 title=(b) Tau vs time,
 xlabel style={ font=\tiny},
 ylabel style={ font=\tiny},
 xlabel={$\tau$},
 ylabel={Query Time(ms)$\times 1000$},
 ymin=.3, ymax=1.2,
 xmin=0.1, xmax=.15,
 legend entries={$\theta=.1$,$\theta=.2$,$\theta=.3$,$\theta=.4$},
 ]
 \addplot+[sharp plot] coordinates
 { (.1,.256) (.12,.363) (.14,.434) (.16,.425) (.18,.421) (.2,.415)};
 \addplot+[sharp plot] coordinates
 {(.1,.267) (.12,.376) (.14,.383) (.16,.435) (.18,.434) (.2,.428)};
 \addplot+[sharp plot] coordinates
 {(.1,.287) (.12,.454) (.14,.443) (.16,.440) (.18,.438) (.2,.433)};
 \addplot+[sharp plot] coordinates
 {(.1,.223) (.12,.492) (.14,.481) (.16,.489) (.18,.459) (.2,.439)};
 \end{axis}
 \end{tikzpicture}
  \begin{tikzpicture}
 \begin{axis}[
 xticklabel style={
 /pgf/number format/precision=2,
 /pgf/number format/fixed,
 /pgf/number format/fixed zerofill,
 },
 xlabel style={ font=\tiny},
 ylabel style={ font=\tiny},
 xlabel=$\tau_{min}$,
 ylabel={Query Time(ms)$\times 1000$},
 legend pos=north east,
 legend style={draw=none, font=\tiny},
 title=(c) Tau(min) vs time,
 ymin=.300, ymax=1.600,
 xmin=0.03, xmax=.2,
 legend entries={$\theta=.1$,$\theta=.2$,$\theta=.3$,$\theta=.4$},
 ]
 \addplot+[sharp plot] coordinates
 {(.04,.856) (.06,.834) (.08,.443) (.1,.495025) (.12,.459) (.14,.449) (.16,.365)};
 \addplot+[sharp plot] coordinates
 {(.04,.970) (.06,.740) (.08,.543) (.1,.485025) (.12,.489) (.14,.455) (.16,.345)};
 \addplot+[sharp plot] coordinates
 {(.04,.986) (.06,.844) (.08,.640) (.1,.550025) (.12,.550) (.14,.469) (.16,.365)};
 \addplot+[sharp plot] coordinates
 {(.04,1.175) (.06,.854) (.08,.743) (.1,.595025) (.12,.559) (.14,.459) (.16,.378)};
 \end{axis}
 \end{tikzpicture}
 \hskip -5pt
   \begin{tikzpicture}
 \begin{axis}[
 xticklabel style={
 /pgf/number format/precision=2,
 /pgf/number format/fixed,
 /pgf/number format/fixed zerofill,
 },
 xlabel style={ font=\tiny},
 ylabel style={ font=\tiny},
 xlabel=$m$,
 ylabel={Query Time(ms)$\times 1000$},
 legend pos=north west,
 legend style={draw=none, font=\tiny},
 title=(d) $m$ vs time,
 ymin=.300, ymax=150,
 xmin=2, xmax=25,
 legend entries={$\theta=.1$,$\theta=.2$,$\theta=.3$,$\theta=.4$},
 ]
 \addplot+[sharp plot] coordinates
 {(3,.526) (6,.97) (9,.483) (12,.595025) (18,24) (21,29) (23,38.5)};
 \addplot+[sharp plot] coordinates
 {(3,.540) (6,.780) (9,.573) (12,.525025) (18,28) (21,35) (23,42.2)};
 \addplot+[sharp plot] coordinates
 {(3,.536) (6,.634) (9,.790) (12,.740025) (18,35) (21,45) (23,48.5)};
 \addplot+[sharp plot] coordinates
 {(3,.575) (6,.654) (9,.713) (12,.855025) (18,45) (21,55) (23,65.6)};
 \end{axis}
 \end{tikzpicture}
   \caption{Substring searching query Time for different string lengths($n$), query threshold value $\tau$, construction time threshold parameter $\tau_{min}$ and query substring length $m$.}
   \label{experimentsubstring}
 \end{figure*}
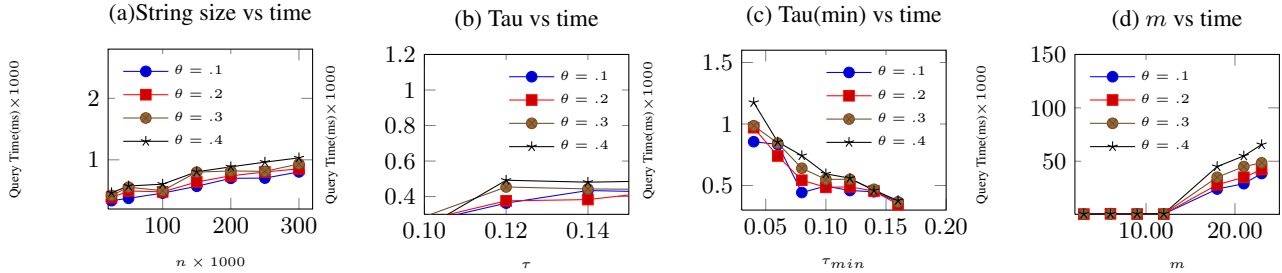
 
 \begin{figure*}[t]
   \label{fig:experiment1}
  \pgfplotsset{width=4.3cm}
   \begin{tikzpicture}
 \begin{axis}[legend pos=north west,
 legend style={draw=none, font=\tiny},
 title=(a)String size vs time,
 xlabel style={ font=\tiny},
 ylabel style={ font=\tiny},
 xlabel={$n\times 1000$},
 ylabel={Query Time(ms)$\times 1000$},
 ymin=.2, ymax=1.3,
 xmin=20, xmax=320,
 legend entries={$\theta=.1$,$\theta=.2$,$\theta=.3$,$\theta=.4$},
 ]
 \addplot+[sharp plot] coordinates
 {(25,.375714) (50,.394308) (100,.398582) (150,.369882) (200,.399356) (250,.402752) (300,.403371)};
 \addplot+[sharp plot] coordinates
 {(25,.425692) (50,.434395) (100,.438703) (150,.439857) (200,.439613) (250,.442163) (300,.453752)};
 \addplot+[sharp plot] coordinates
 {(25,.430305) (50,.454604) (100,.478645) (150,.499436) (200,.519623) (250,.512354) (300,.533775)};
 \addplot+[sharp plot] coordinates
 {(25,.435758) (50,.474108) (100,.484230) (150,.505612) (200,.535623) (250,.562143) (300,.593156)};
 \end{axis}
 \end{tikzpicture}
 \hskip -5pt
  \begin{tikzpicture}
 \begin{axis}[
 xticklabel style={
 /pgf/number format/precision=2,
 /pgf/number format/fixed,
 /pgf/number format/fixed zerofill,
 },
 legend pos=north east,
 legend style={draw=none, font=\tiny},
 title=(b) Tau vs time,
 xlabel style={ font=\tiny},
 ylabel style={ font=\tiny},
 xlabel={$\tau$},
 ylabel={Query Time(ms)$\times 1000$},
 ymin=.3, ymax=.7,
 xmin=0.1, xmax=.15,
 legend entries={$\theta=.1$,$\theta=.2$,$\theta=.3$,$\theta=.4$},
 ]
 \addplot+[sharp plot] coordinates
 { (.1,.356) (.12,.333) (.14,.334) (.16,.325) (.18,.321) (.2,.315)};
 \addplot+[sharp plot] coordinates
 {(.1,.367) (.12,.356) (.14,.343) (.16,.335) (.18,.334) (.2,.328)};
 \addplot+[sharp plot] coordinates
 {(.1,.387) (.12,.354) (.14,.343) (.16,.340) (.18,.338) (.2,.333)};
 \addplot+[sharp plot] coordinates
 {(.1,.423) (.12,.412) (.14,.401) (.16,.389) (.18,.359) (.2,.339)};
 \end{axis}
 \end{tikzpicture}
  \begin{tikzpicture}
 \begin{axis}[
 xticklabel style={
 /pgf/number format/precision=2,
 /pgf/number format/fixed,
 /pgf/number format/fixed zerofill,
 },
 xlabel style={ font=\tiny},
 ylabel style={ font=\tiny},
 xlabel=$\tau_{min}$,
 ylabel={Query Time(ms)$\times 1000$},
 legend pos=north east,
 legend style={draw=none, font=\tiny},
 title=(c) Tau(min) vs time,
 ymin=.300, ymax=1.300,
 xmin=0.03, xmax=.2,
 legend entries={$\theta=.1$,$\theta=.2$,$\theta=.3$,$\theta=.4$},
 ]
 \addplot+[sharp plot] coordinates
 {(.04,.826) (.06,.814) (.08,.343) (.1,.395025) (.12,.359) (.14,.349) (.16,.355)};
 \addplot+[sharp plot] coordinates
 {(.04,.940) (.06,.730) (.08,.443) (.1,.385025) (.12,.389) (.14,.355) (.16,.345)};
 \addplot+[sharp plot] coordinates
 {(.04,.936) (.06,.874) (.08,.640) (.1,.450025) (.12,.450) (.14,.369) (.16,.365)};
 \addplot+[sharp plot] coordinates
 {(.04,.975) (.06,.884) (.08,.643) (.1,.495025) (.12,.459) (.14,.359) (.16,.358)};
 \end{axis}
 \end{tikzpicture}
 \hskip -5pt
   \begin{tikzpicture}
 \begin{axis}[
 xticklabel style={
 /pgf/number format/precision=2,
 /pgf/number format/fixed,
 /pgf/number format/fixed zerofill,
 },
 xlabel style={ font=\tiny},
 ylabel style={ font=\tiny},
 xlabel=$m$,
 ylabel={Query Time(ms)$\times 1000$},
 legend pos=north west,
 legend style={draw=none, font=\tiny},
 title=(d) $m$ vs time,
 ymin=.300, ymax=100,
 xmin=2, xmax=25,
 legend entries={$\theta=.1$,$\theta=.2$,$\theta=.3$,$\theta=.4$},
 ]
 \addplot+[sharp plot] coordinates
 {(3,.426) (6,.94) (9,.443) (12,.495025) (18,22) (21,25) (23,28.5)};
 \addplot+[sharp plot] coordinates
 {(3,.440) (6,.530) (9,.543) (12,.575025) (18,26) (21,29) (23,30.2)};
 \addplot+[sharp plot] coordinates
 {(3,.436) (6,.674) (9,.740) (12,.780025) (18,32) (21,35) (23,38.5)};
 \addplot+[sharp plot] coordinates
 {(3,.475) (6,.684) (9,.743) (12,.895025) (18,42) (21,45) (23,48.6)};
 \end{axis}
 \end{tikzpicture}
   \caption{String listing query Time for different string lengths($n$), query threshold value $\tau$, construction time threshold parameter $\tau_{min}$ and query substring length $m$.}
   \label{experimentstringlisting}
 \end{figure*}
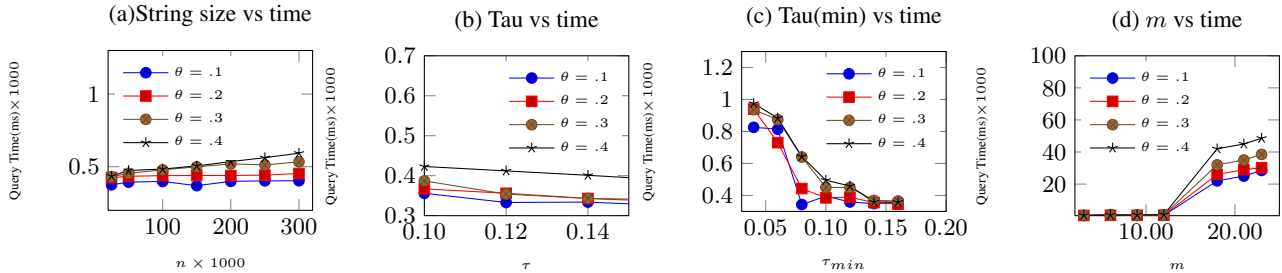

In this section we evaluate the performance of our substring searching and string listing index. We use a collection of query substrings and observe the effect of varying the key parameters. Our experiments show that, for short query substrings, uncertain string length does not affect the query performance. For long query substrings, our index fails to achieve optimal query time. However this does not deteriorate the average query time by big margin, since the probability of match also decreases significantly as substring gets longer. Index construction time is proportional to uncertain string size and probability threshold parameter $\tau_{min}$.

We have implemented the proposed indexing scheme in C++. The experiments are performed on a $64$-bit machine with an Intel Core i5 CPU 3.33GHz processor and $8$GB RAM running Ubuntu. We present experiments along with analysis of performance.
\subsection{Dataset}
We use a synthetic datasets obtained from their real counterparts. We use a concatenated protein sequence of mouse and human (alphabet size $|\Sigma | = 22$), and break it arbitrarily into shorter strings. For each string $s$ in the dataset we first obtain a set $A(s)$ of strings that are within edit distance 4 to $s$. Then a character-level probabilistic string $S$ for string $s$ is generated such that, for a position $i$, the pdf of $S[i]$ is based on the normalized frequencies of the letters in the $i$-th position
of all the strings in $A(s)$. We denote by $\theta$ the fraction of uncertain characters in the string. $\theta$ is varied between $0.1$ to $0.5$ to generate strings with different degree of uncertainty. The string length distributions in this dataset follow approximately a normal distribution in the range of $[20, 45]$. The average number of choices that each probabilistic character $S[i]$ may have is set to $5$.

\subsection{Query Time for Different String Lengths($n$) and Fraction of Uncertainty($\theta$)}
We evaluate the query time for different string lengths $n$, ranging from $2K$ to $300K$ and $\theta$ ranging from 0.1 to 0.5. Figure ~\ref{experimentsubstring}(a) and Figure ~\ref{experimentstringlisting}(a), shows the query times for substring searching and string listing. Note that, $n$ is number of positions in the uncertain string where each position can have multiple characters. We take the average time for query lengths of {10,100,500,1000}. We use $\tau_{min}=0.1$ and query threshold $\tau=0.2$.  As shown in the figures, query times does not show much irregularity in performance when the length of string goes high. This is because for shorter query length, our index achieves optimal query time. Although for longer queries, our index achieves $O(m\times occ)$ time, longer query strings probability of occurrence gets low as string grows longer resulting in less number of outputs. However when fraction of uncertainty($\theta$) increases in the string, performance shows slight decrease as query time increases slightly. This is because longer query strings are more probable to match with strings with high level of uncertainty.

\subsection{Query Time for Different $\tau$ and Fraction of Uncertainty($\theta$)}
In Figure~\ref{experimentsubstring}(b) and Figure~\ref{experimentstringlisting}(b), we show the average query times for string matching and string listing for probability threshold $\tau=0.04,0.06,0.08,0.1,0.12$ for fixed $\tau_{min}=0.1$. In terms of performance, query time increases with decreasing $\tau$. This is because more matching is probable for smaller $\tau$. Larger $\tau$ reduces the output size, effectively reducing the query time as well.

\subsection{Query Time for Different $\tau_{min}$ and Fraction of Uncertainty($\theta$)}
In Figure~\ref{experimentsubstring}(c) and Figure~\ref{experimentstringlisting}(c), we show the average query times for string matching and string listing for probability threshold $\tau_{min}=0.04,0.06,0.08,0.1,0.12$ which shows slight impact of $\tau_{min}$ over query time. 
     
\subsection{Query Time for Different Substring Length $m$ and Fraction of Uncertainty($\theta$)}
In figure ~\ref{experimentsubstring}(d) and figure Figure~\ref{experimentstringlisting}(d), we show the average query times for string matching and string listing. As it can be seen long pattern length drastically increases the query time. 

\subsection{Construction Time for Different String \\Lengths and Fraction of Uncertainty($\theta$)}
\begin{figure*}
   \label{fig:experiments}
  \pgfplotsset{width=4.8cm}

  \begin{tikzpicture}
 \begin{axis}[
 title=(a)String size vs time,
 xlabel={$n\times 100$},
 ylabel={Construction Time(Seconds)},
 legend pos=north west,
 legend style={draw=none, font=\tiny},
 ymin=20, ymax=1200,
 xmin=20, xmax=300,
 legend entries={$\theta=.1$,$\theta=.2$,$\theta=.3$,$\theta=.4$},
 ]
 \addplot+[sharp plot] coordinates
 {(25,45) (50,80) (100,150) (150,287) (200,425) (250,472) (300,510)};
 \addplot+[sharp plot] coordinates
 {(25,46) (50,88) (100,180) (150,327) (200,428) (250,530) (300,555)};
 \addplot+[sharp plot] coordinates
 {(25,55) (50,93) (100,196) (150,374) (200,465) (250,550) (300,655)};
 \addplot+[sharp plot] coordinates
 {(25,68) (50,98) (100,220) (150,425) (200,538) (250,650) (300,735)};
 \end{axis}
 \end{tikzpicture}
 \hskip 5pt
    \begin{tikzpicture}
 \begin{axis}[
 title=(a)$\tau_{min}$ vs time,
 xlabel={$n\times 100$},
 ylabel={Construction Time(Seconds)},
 legend pos=north east,
 legend style={draw=none, font=\tiny},
 ymin=20, ymax=1500,
 xmin=0.05, xmax=0.2,
 legend entries={$\theta=.1$,$\theta=.2$,$\theta=.3$,$\theta=.4$},
 ]
 \addplot+[sharp plot] coordinates
 {(.04,810) (.06,650) (.08,580) (.1,400) (.12,325) (.14,272) (.16,110)};
 \addplot+[sharp plot] coordinates
 {(.04,865) (.06,690) (.08,630) (.1,450) (.12,328) (.14,280) (.16,115)};
 \addplot+[sharp plot] coordinates
 {(.04,925) (.06,780) (.08,660) (.1,490) (.12,335) (.14,290) (.16,125)};
 \addplot+[sharp plot] coordinates
 {(.04,995) (.06,810) (.08,780) (.1,550) (.12,338) (.14,295) (.16,165)};
 \end{axis}
 \end{tikzpicture}
 \hskip 5pt
   \begin{tikzpicture}
 \begin{axis}[
 title=(a)String size vs index space,
 xlabel={$n\times 100$},
 ylabel={Index space (MB)},
 legend pos=north west,
 legend style={draw=none, font=\tiny},
 ymin=20, ymax=1200,
 xmin=20, xmax=300,
 legend entries={$\theta=.1$,$\theta=.2$,$\theta=.3$,$\theta=.4$},
 ]
 \addplot+[sharp plot] coordinates
 {(25,46.095) (50,101.22) (100,165.27) (150,325.185) (200,576.595) (250,526.595) (300,588.595)};

 \addplot+[sharp plot] coordinates
 {(25,48) (50,104) (100,169) (150,328) (200,481) (250,629) (300,623)};

 \addplot+[sharp plot] coordinates
 {(25,50) (50,106) (100,172) (150,335) (200,486) (250,708) (300,697)};

 \addplot+[sharp plot] coordinates
 {(25,55) (50,109) (100,175) (150,346) (200,496) (250,749) (300,802)};
 \end{axis}
 \end{tikzpicture}
   \caption{Construction time and index space for different string lengths($n$) and probability threshold $\tau_{min}=.1$}
   \label{experimentconstruction}
 \end{figure*}
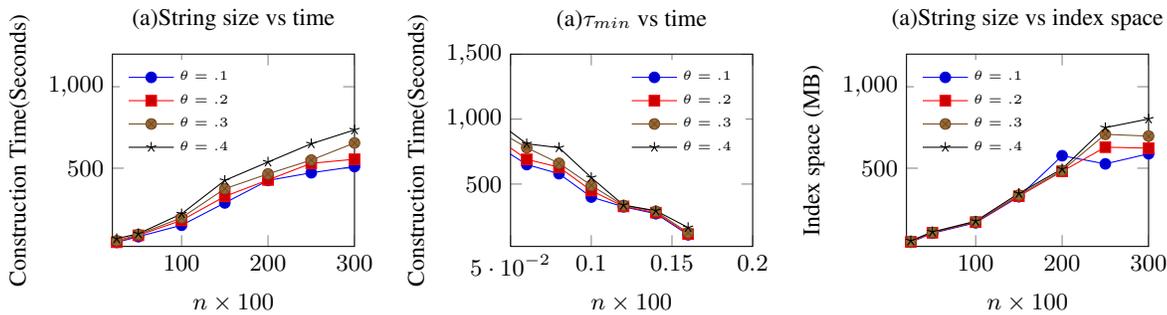

Figure~\ref{experimentconstruction}(a) shows the index construction times for uncertain string length $n$ ranging from $2K$ to $300K$. We can see that the construction time is proportional to the string length $n$. Increasing uncertainty factor $\theta$ also impacts the construction time as more permutation is possible with increasing uncertain positions. Figure~\ref{experimentconstruction}(b) shows the impact of $\theta$ on construction time.

\subsection{Space Usage}
Theoretical bound for our index is $O(n)$. However, this bound can have hidden multiplicative constant. Here we elaborate more on the actual space used for our index.

For our indexes, we construct the regular string $t$ of length $N=O((\frac{1}{\tau_{min}})^2n)$ by concatenating all the extended maximal factors based on threshold $\tau_{min}$. We do not store the string $t$ in our index. We built RMQ structures $RMQ_i$ for $i=1,\dots ,\log n$ which takes $O(N\log n)$ bits. The practical space usage of RMQ is usually very small with hidden multiplicative constant of $2-3$. So the average space usage of our RMQ structure in total can be stated as $3N$ words. For a query string $p$, we find the suffix range of $p$ in the concatenated extended maximum factor string $t$. For this purpose, instead of using Generalized Suffix Tree(GST), we use its space efficient version i.e., a compressed suffix array $(CSA)$ of $t$. There are many versions of CSA's available in literature. For our purpose we use the one in~\cite{BelazzouguiN11} that occupies $N\log\sigma+o(N\log\sigma)+O(N)$ bits space and retrieves the suffix range of query string $p$ in $O(p)$ time. In practice, this structure takes about $2.5N$ words space. We also store an array $D$ of size $N$ storing the partial probabilities, which takes approximately $4N$ bytes of space. Finally $Pos$ array is used for position transformation, taking $N$ words space. Summing up all the space usage, our index takes approximately $3N+2.5N+4N+N = 10.5N = (\frac{1}{\tau_{min}})^2 10.5n$. Figure Figure~\ref{experimentconstruction}(c) shows the space usage for different string length($n$) and $\theta$.

\section{Conclusions}
\label{conclusion}
In this paper we presented indexing framework for searching in uncertain strings. We tackled the problem of searching a deterministic substring in uncertain string and proposed both exact and approximate solution. We also formulated the uncertain string listing problem and proposed index for string listing from a uncertain string collection. Our indexes can support arbitrary values of probability threshold parameter. Uncertain string searching is still largely unexplored area. Constructing more efficient index, variations of the string searching problem satisfying diverse query constraints are some interesting future work direction.

\balance


\bibliographystyle{abbrv}
\bibliography{uncertain}  

\begin{thebibliography}{10}

\bibitem{AmirCIKZ08}
A.~Amir, E.~Chencinski, C.~S. Iliopoulos, T.~Kopelowitz, and H.~Zhang.
\newblock Property matching and weighted matching.
\newblock {\em Theor. Comput. Sci.}, 395(2-3):298--310, 2008.

\bibitem{BelazzouguiN11}
D.~Belazzougui and G.~Navarro.
\newblock Alphabet-independent compressed text indexing.
\newblock In {\em ESA}, pages 748--759, 2011.

\bibitem{BerneckerKRVZ12}
T.~Bernecker, H.~Kriegel, M.~Renz, F.~Verhein, and A.~Z{\"{u}}fle.
\newblock Probabilistic frequent pattern growth for itemset mining in uncertain
  databases.
\newblock In {\em Scientific and Statistical Database Management - 24th
  International Conference, {SSDBM} 2012, Chania, Crete, Greece, June 25-27,
  2012. Proceedings}, pages 38--55, 2012.

\bibitem{ChaudhuriGK06}
S.~Chaudhuri, V.~Ganti, and R.~Kaushik.
\newblock A primitive operator for similarity joins in data cleaning.
\newblock In {\em Proceedings of the 22nd International Conference on Data
  Engineering, {ICDE} 2006, 3-8 April 2006, Atlanta, GA, {USA}}, page~5, 2006.

\bibitem{cheng2004efficient}
R.~Cheng, Y.~Xia, S.~Prabhakar, R.~Shah, and J.~S. Vitter.
\newblock Efficient indexing methods for probabilistic threshold queries over
  uncertain data.
\newblock In {\em Proceedings of the Thirtieth international conference on Very
  large data bases-Volume 30}, pages 876--887. VLDB Endowment, 2004.

\bibitem{ChuiK08}
C.~K. Chui and B.~Kao.
\newblock A decremental approach for mining frequent itemsets from uncertain
  data.
\newblock In {\em Advances in Knowledge Discovery and Data Mining, 12th
  Pacific-Asia Conference, {PAKDD} 2008, Osaka, Japan, May 20-23, 2008
  Proceedings}, pages 64--75, 2008.

\bibitem{ChuiKH07}
C.~K. Chui, B.~Kao, and E.~Hung.
\newblock Mining frequent itemsets from uncertain data.
\newblock In {\em Advances in Knowledge Discovery and Data Mining, 11th
  Pacific-Asia Conference, {PAKDD} 2007, Nanjing, China, May 22-25, 2007,
  Proceedings}, pages 47--58, 2007.

\bibitem{dalvi2007efficient}
N.~Dalvi and D.~Suciu.
\newblock Efficient query evaluation on probabilistic databases.
\newblock {\em The VLDB Journal}, 16(4):523--544, 2007.

\bibitem{dash2009automatic}
S.~Dash, K.~Chon, S.~Lu, and E.~Raeder.
\newblock Automatic real time detection of atrial fibrillation.
\newblock {\em Annals of biomedical engineering}, 37(9):1701--1709, 2009.

\bibitem{FHS07}
J.~Fischer and V.~Heun.
\newblock {A New Succinct Representation of RMQ-Information and Improvements in
  the Enhanced Suffix Array}.
\newblock In {\em ESCAPE}, pages 459--470, 2007.

\bibitem{FHS08}
J.~Fischer, V.~Heun, and H.~M. St{\"u}hler.
\newblock {Practical Entropy-Bounded Schemes for $O(1)$-Range Minimum Queries}.
\newblock In {\em IEEE DCC}, pages 272--281, 2008.

\bibitem{GeL11}
T.~Ge and Z.~Li.
\newblock Approximate substring matching over uncertain strings.
\newblock {\em {PVLDB}}, 4(11):772--782, 2011.

\bibitem{GravanoIJKMS01}
L.~Gravano, P.~G. Ipeirotis, H.~V. Jagadish, N.~Koudas, S.~Muthukrishnan, and
  D.~Srivastava.
\newblock Approximate string joins in a database (almost) for free.
\newblock In {\em {VLDB} 2001, Proceedings of 27th International Conference on
  Very Large Data Bases, September 11-14, 2001, Roma, Italy}, pages 491--500,
  2001.

\bibitem{hon2009space}
W.-K. Hon, R.~Shah, and J.~S. Vitter.
\newblock Space-efficient framework for top-k string retrieval problems.
\newblock In {\em Foundations of Computer Science, 2009. FOCS'09. 50th Annual
  IEEE Symposium on}, pages 713--722. IEEE, 2009.

\bibitem{JestesLYY10}
J.~Jestes, F.~Li, Z.~Yan, and K.~Yi.
\newblock Probabilistic string similarity joins.
\newblock In {\em Proceedings of the {ACM} {SIGMOD} International Conference on
  Management of Data, {SIGMOD} 2010, Indianapolis, Indiana, USA, June 6-10,
  2010}, pages 327--338, 2010.

\bibitem{kanagal2009indexing}
B.~Kanagal and A.~Deshpande.
\newblock Indexing correlated probabilistic databases.
\newblock In {\em Proceedings of the 2009 ACM SIGMOD International Conference
  on Management of data}, pages 455--468. ACM, 2009.

\bibitem{LeungH09}
C.~K. Leung and B.~Hao.
\newblock Mining of frequent itemsets from streams of uncertain data.
\newblock In {\em Proceedings of the 25th International Conference on Data
  Engineering, {ICDE} 2009, March 29 2009 - April 2 2009, Shanghai, China},
  pages 1663--1670, 2009.

\bibitem{LiLL08}
C.~Li, J.~Lu, and Y.~Lu.
\newblock Efficient merging and filtering algorithms for approximate string
  searches.
\newblock In {\em Proceedings of the 24th International Conference on Data
  Engineering, {ICDE} 2008, April 7-12, 2008, Canc{\'{u}}n, M{\'{e}}xico},
  pages 257--266, 2008.

\bibitem{li2011unified}
J.~Li, B.~Saha, and A.~Deshpande.
\newblock A unified approach to ranking in probabilistic databases.
\newblock {\em The VLDB Journal—The International Journal on Very Large Data
  Bases}, 20(2):249--275, 2011.

\bibitem{LiBKP14}
Y.~Li, J.~Bailey, L.~Kulik, and J.~Pei.
\newblock Efficient matching of substrings in uncertain sequences.
\newblock In {\em Proceedings of the 2014 {SIAM} International Conference on
  Data Mining, Philadelphia, Pennsylvania, USA, April 24-26, 2014}, pages
  767--775, 2014.

\bibitem{lilley1996nomenclature}
D.~M. Lilley, R.~M. Clegg, S.~Diekmann, N.~C. Seeman, E.~Von~Kitzing, and P.~J.
  Hagerman.
\newblock Nomenclature committee of the international union of biochemistry and
  molecular biology (nc- iubmb) a nomenclature of junctions and branchpoints in
  nucleic acids recommendations 1994.
\newblock {\em European Journal of Biochemistry, s. FEBS J}, 230(1):1--2, 1996.

\bibitem{st2}
E.~M. McCreight.
\newblock A space-economical suffix tree construction algorithm.
\newblock {\em J. ACM}, 23(2):262--272, 1976.

\bibitem{Navarro01}
G.~Navarro.
\newblock A guided tour to approximate string matching.
\newblock {\em {ACM} Comput. Surv.}, 33(1):31--88, 2001.

\bibitem{patil2014similarity}
M.~Patil and R.~Shah.
\newblock Similarity joins for uncertain strings.
\newblock In {\em Proceedings of the 2014 ACM SIGMOD international conference
  on Management of data}, pages 1471--1482. ACM, 2014.

\bibitem{re2007efficient}
C.~Re, N.~Dalvi, and D.~Suciu.
\newblock Efficient top-k query evaluation on probabilistic data.
\newblock In {\em Data Engineering, 2007. ICDE 2007. IEEE 23rd International
  Conference on}, pages 886--895. IEEE, 2007.

\bibitem{singh2007indexing}
S.~Singh, C.~Mayfield, S.~Prabhakar, R.~Shah, and S.~Hambrusch.
\newblock Indexing uncertain categorical data.
\newblock In {\em Data Engineering, 2007. ICDE 2007. IEEE 23rd International
  Conference on}, pages 616--625. IEEE, 2007.

\bibitem{tao2005indexing}
Y.~Tao, R.~Cheng, X.~Xiao, W.~K. Ngai, B.~Kao, and S.~Prabhakar.
\newblock Indexing multi-dimensional uncertain data with arbitrary probability
  density functions.
\newblock In {\em Proceedings of the 31st international conference on Very
  large data bases}, pages 922--933. VLDB Endowment, 2005.

\bibitem{st1}
P.~Weiner.
\newblock Linear pattern matching algorithms.
\newblock In {\em SWAT (FOCS)}, pages 1--11, 1973.

\end{thebibliography}
%


%
\pagebreak
\begin{samepage}
\onecolumn
\begin{appendix}
\section{Algorithm for Substring Searching in General Uncertain String}
\begin{algorithm}

\SetKwInOut{Input}{input}
\SetKwInOut{Output}{output}
 \Input{A general uncertain string $S$, probability threshold $\tau_{min}$}
 \Output{Suffix tree over $t$, suffix array $A$, Position transformation array $Pos$, successive multiplicative probability array $C$, $RMQ_i,i=1,\dots,\log n$}
 Transform $S$ into special uncertain string $X$ for $\tau_{min}$ using lemma~\ref{Extended}\\
 Build position transformation array $Pos$\\
 Build deterministic string $t$ from $X$\\
 Build suffix tree over $t$\\
 Build suffix array $A$ over $t$\\
 \tcp{Building successive multiplicative probability array}
 $C[1]=Pr(c_1^1)$\\
 \For{$i=2;i\leq n; i++$}{
 	$C[i]=C[i-1]\times Pr(c_i^i)$ 
 }
 \tcp{Building $C_i,i=1,\dots ,\log n$ arrays}
 \For{$i=1;i\leq \log n; i++$}{
 	\For{$j=1;j\leq n; j++$}{
 		$C_i[j]=C[A[j]+i-1]/C[A[j]-1]$
 	}
 }
 \tcp{Duplicate elimination in $C_i$}
 \For{$i=1;i\leq \log n; i++$}{
 	Find the set of locus nodes $L_i$ in the suffix tree
 	Compute the set of suffix ranges corresponding to $L_i$
 	Use $Pos$ array for duplicate elimination in $C_i$ for each range
 }	 
 \For{$i=1;i\leq \log n; i++$}{
 	Build $RMQ_i$ over the array $C_i$
 }

 \caption{Algorithm General-Short-Substring-Index-Construction}
 \label{alg_generalCon}
\end{algorithm}

\begin{algorithm}
\label{alg_generalQuery}
\SetKwInOut{Input}{input}
\SetKwInOut{Output}{output}
 \Input{Query substring $p$, probability threshold $\tau\geq \tau_{min}$}
 \Output{Occurrence positions of $p$ in $X$ with probability of occurrence greater than $\tau$}
 $m=length(p)$\\
 call RecursiveRmq($m,1,n$)\\

 \begin{algorithmic}[]
\Function{RecursiveRmq}{$i,l,r$}\Comment{Recursive RMQ method}
	\State $max=RMQ_m(l,r)$
	\State $max'=A[max]$
	\State \If{$C[max'+i]/C[max']>\tau$}{
	  \State Output $Pos[max']$
      \State Call $RecursiveRmq(m,l,max-1)$
      \State Call $RecursiveRmq(m,max+1,r)$
      }
    \EndFunction
\end{algorithmic}

 \caption{Algorithm General-Short-Substring-Query-Answering}
 \label{alg_generalQuery}
\end{algorithm}

\pagebreak

\section{Running Example of Algorithm 4}

\hfill \break

\begin{figure}[h!]

\begin{subfigure}{0.4\textwidth}
	\normalsize
    \begin{center}
    \begin{tabular}{|c|c|c|c|}
    \hline
    S[1] & S[2] & S[3] & S[4]\\ \hline
    Q .7 &Q .3 &P 1 &A .4\\ \hline
    S .3 &P .7 &    &F .3\\ \hline
         &     &    &P .2\\\hline
    	 &     &    &Q .1\\
\hline
    \end{tabular}
    \end{center}
    \caption{General uncertain string $S$}

\end{subfigure}

\hfill \break

\begin{subfigure}{0.4\textwidth}
   \tiny
   \setlength{\tabcolsep}{3pt}
   \begin{center}
       \begin{tabular}{|cccccccccccccccccccccccccccccccccccccccccccccc|}
    \hline
   \large t:\small &  Q  &Q  &P  &\$  &Q  &P  &P  &A  &\$  &Q  &P  &P  &F  &\$  &Q  &P  &A  &\$  &Q  &P  &F  &\$  &T  &P  &A  &\$ &T &P &F &\$ &P &A &\$&P&F&\$&P&P&\$&A&\$&F&\$&P&\$ \\ \hline
   \large Pos:\small & 1  &2  &3  &\$  &1  &2  &3  &4  &\$  &1  &2  &3  &4  &\$  &2  &3  &4  &\$  &2  &3  &4  &\$  &2  &3  &4  &\$ &2 &3 &4 &\$ &3 &4 &\$&3&4&\$&3&4&\$&4&\$&4&\$&4&\$ \\ \hline
    \large C:\small & .7 &.21 &.21&-1  &.7 &.49&.49&.19&-1  &.7 &.49&.49&.14&-1  &.5 &.5 &.2 &-1 &.5 &.5 &.15 &-1 &.5 &.5 &.2 &-1 &.5 &.5 &.15 &-1 &1 &.4 &-1&1&.3&-1&1&.2&-1&.4&-1&.3&-1&.2&-1 \\ \hline

    \end{tabular}
    \end{center}
    \caption{Deterministic string $t$, position transformation array $Pos$, successive multiplicative probability array $C$.}
\end{subfigure}

\hfill \break
\hfill \break
\begin{subfigure}{0.6\textwidth}
      \includegraphics[width=0.92\textwidth]{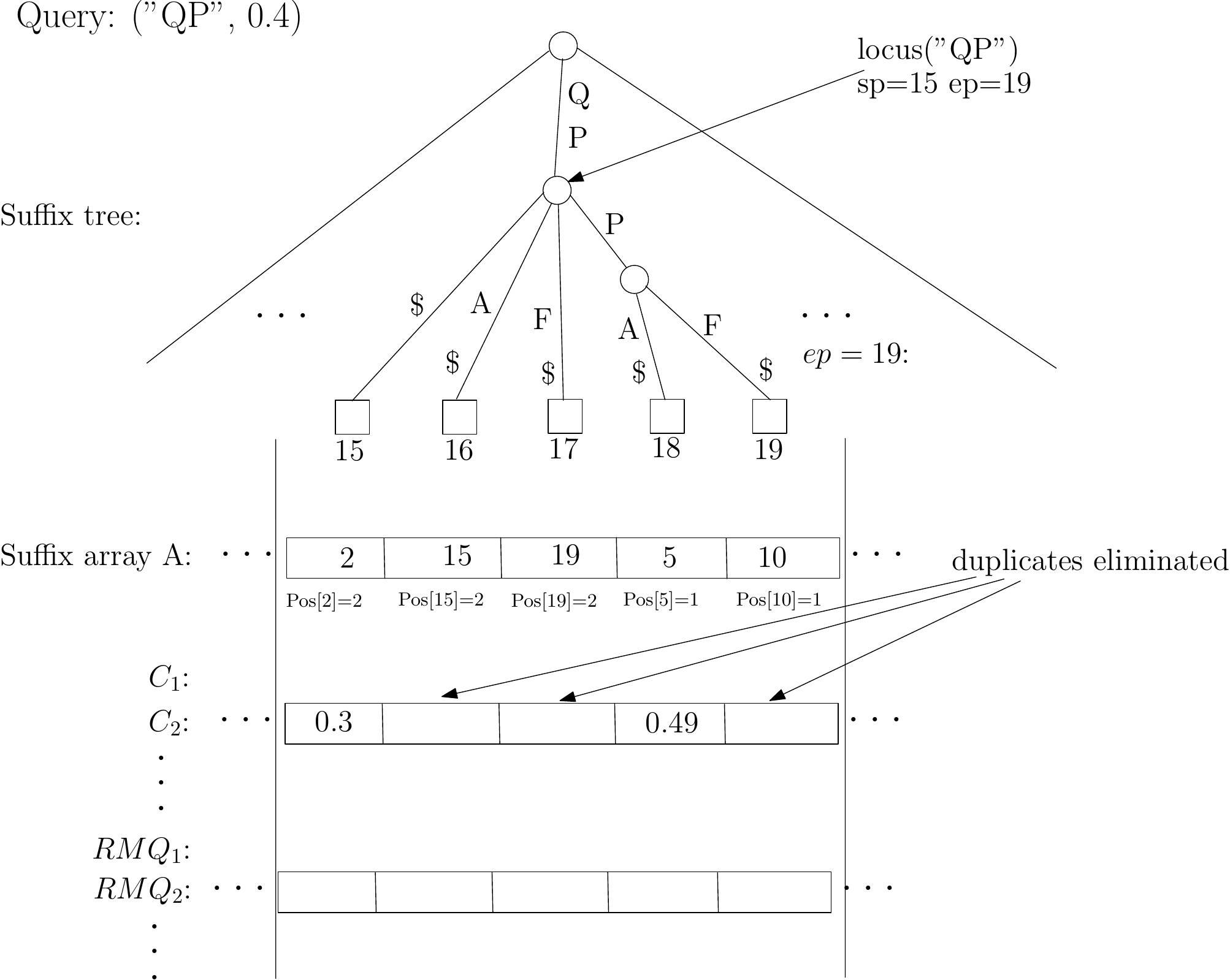}
\end{subfigure}
    \begin{subfigure}{0.39\textwidth}
      \includegraphics[width=0.7\textwidth]{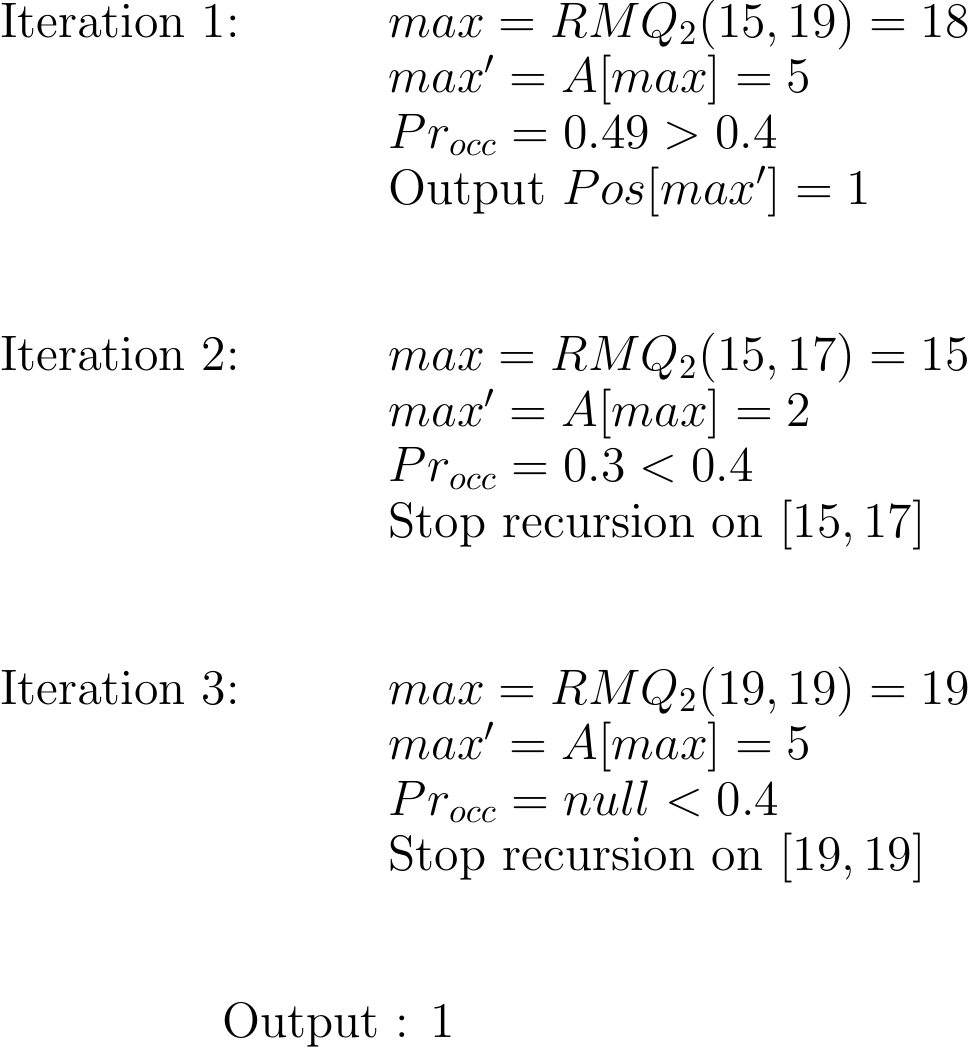}
\end{subfigure}
 \caption{Running example of Algorithm~\ref{alg_generalQuery}}
\label{queryappendix}
\end{figure}
\end{appendix}
\end{samepage}

\end{document}